\documentclass[rmp,aps,amsfonts,amsmath,amssymb,nofootinbib,twocolumn]{revtex4}
\usepackage{graphics}
\usepackage{hyperref}
\usepackage{epsfig}
\usepackage{color}
\usepackage{bm}

\begin{document}

\title{Low Energy Electrodynamics of Novel Spin Excitations in the Quantum Spin Ice Yb$_2$Ti$_2$O$_7$ }



\author{LiDong Pan$^1$, Se Kwon Kim$^1$, A. Ghosh$^1$, Christopher M. Morris$^1$, Kate A. Ross$^{1,2}$, Edwin Kermarrec$^3$, Bruce D. Gaulin$^{3,4,5}$, S. M. Koohpayeh$^1$, Oleg Tchernyshyov$^1$, and N. P. Armitage$^1$\\
1, Institute for Quantum Matter, Department of Physics and Astronomy, Johns Hopkins University, Baltimore, Maryland 21218, USA
2, NIST Center for Neutron Research, NIST, Gaithersburg, Maryland 20899, USA
3, Department of Physics and Astronomy, McMaster University, Hamilton, Ontario L8S 4M1, Canada
4, Brockhouse Institute for Materials Research, McMaster University, Hamilton, Ontario, L8S 4M1, Canada 
5, Canadian Institute for Advanced Research, 180 Dundas St. W, Toronto, ON, M5G 1Z8, Canada}

\date{\today}


\maketitle

\textbf{In condensed matter systems, the formation of long range order (LRO) with broken symmetry is often accompanied by new types of excitations. However, in many magnetic pyrochlore oxides, geometrical frustration suppresses conventional LRO while at the same time non-trivial spin correlations are observed. For such materials, a natural question to ask then is what is the nature of the excitations in this highly correlated state without broken symmetry?   Frequently the application of a symmetry breaking field can stabilize excitations whose properties still reflect certain aspects of the anomalous state without long-range order.  Here we report evidence of novel magnetic excitations in the quantum spin ice material Yb$_2$Ti$_2$O$_7$, obtained from time-domain terahertz spectroscopy (TDTS).  At large fields, both magnon and two-magnon-like excitations are observed in a $<$001$>$ directed magnetic field illustrating the stabilization of a field induced LRO state. The unique ability of TDTS to measure complex response functions allows a direct study of magnetic responses in different polarization channels, revealing the existence of an unusual \textit{left}-hand polarized magnon. The g-factors of these excitations are dramatically enhanced in the low-field limit, showing a cross-over of these one- and two-magnon states into features consistent with quantum string-like excitations proposed to exist in quantum spin ice in a small $<$001$>$ applied field.}

\bigskip

Geometrical frustration suppresses the formation of conventional long range order (LRO) in magnetic materials to a temperature ($T_C$) much lower than the Curie-Weiss temperature $\Theta_{CW}$ $^1$. This provides the possibility of realizing novel states and excitations, since non-trivial spin correlations are often observed in these materials in the range T$\ll\Theta_{CW}$ even in the absence of LRO$^2$. Magnetic pyrochlore oxides, in which magnetic rare earth ions sit at the vertices of corner-sharing tetrahedra, provide a fascinating material system for the search of such novel quantum ground states and excitations$^3$.

\begin{figure}[t]
\includegraphics[trim = 30 30 60 250,width=7.8cm]{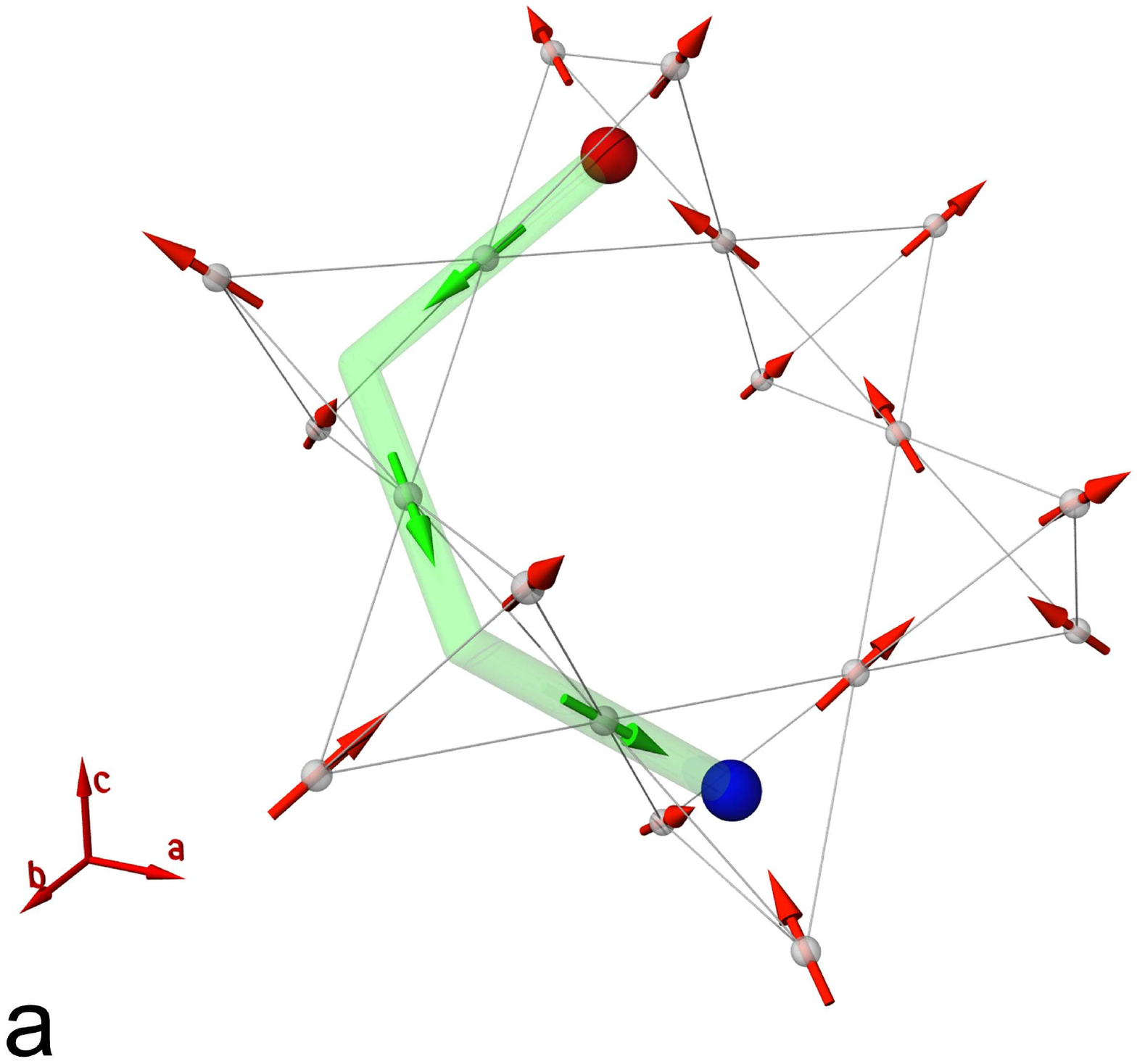}
\includegraphics[trim = 30 30 60 500,width=7.8cm]{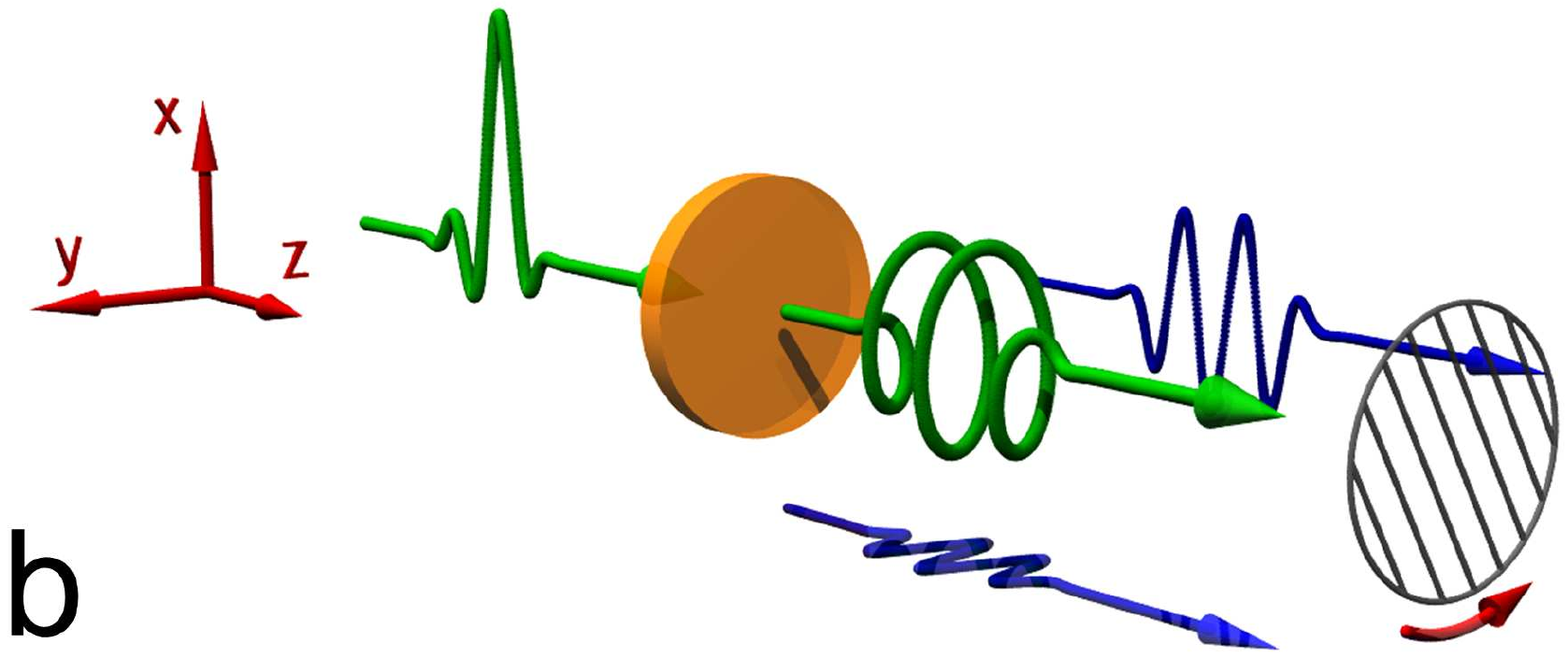}
\label{fig1}
\caption{\textsf{\textbf{Schematic of spin-ice and experimental setup}}  (a) Schematic illustration of a quantum string in $<$001$>$ field in a pyrochlore spin ice. Monopole - antimonopole pairs are shown as red-blue spheres, spins with a negative ${z}$ component are shown as green arrows. Magnetic field is applied along the \textbf{c} direction. (b) Schematic illustration of the experimental geometry. After a linearly polarized incident pulse passes through the sample, the two orthogonal components ($E_X$ and $E_Y$, shown as blue pulses) of the elliptically polarized transmitted pulse are measured with the polarimeter. The rotating wire grid polarizer shown in the figure is a key component of the polarimeter.}
\end{figure}

In spin ice materials such as Dy$_2$Ti$_2$O$_7$ and Ho$_2$Ti$_2$O$_7$, strong crystal field anisotropy forces the spins to point along the local $<$111$>$ direction while long-range dipolar interaction between the spins provides an effective ferromagnetic interaction, giving a situation much resembling proton disorder in common water ice. Enforcing the ices rules in spin ice results in a ground state with each tetrahedron having the so-called ``two-in, two-out" configuration$^{4,5}$. Flipping a single spin creates a pair of magnetic monopoles out of the ground state vacuum in neighboring tetrahedra, which  can subsequently move away from each other through the network$^{6,7}$.   In applied field, a pair of magnetic monopoles with opposite magnetic charges are connected by a string of spins aligned against the field (Fig. 1(a)). Recent neutron scattering experiments have shown evidence for the existence of thermally activated Dirac strings and magnetic monopoles in classical spin ice materials$^{8,9}$.

Recently the material Yb$_2$Ti$_2$O$_7$ has attracted considerable attention$^{10-21}$. Although the exact form of the low temperature ground state is still under debate, the nature of the magnetic interactions place Yb$_2$Ti$_2$O$_7$ in the \textit{quantum spin ice} regime. High-field inelastic neutron scattering (INS) experiments have obtained the exchange interaction parameters at low temperature, which demonstrates that in this material the ferromagnetic Ising type exchange interaction is larger than the other terms in the Hamiltonian that lead to considerable quantum fluctuations and dynamics$^{14}$.  From considerations of crystal-field levels, its low-energy spin sector may be reduced to that of an effective spin-1/2 moment.  A number of interesting effects have been proposed for this material, including it being a quantum spin liquid, whose elementary excitations carry fractional quantum numbers. Such a state supports an emergent quantum electrodynamics with a photon mode at low energy$^{22-24}$.  It was proposed recently that under weak applied  $<$001$>$ magnetic field (or in the presence of spontaneous magnetization), the elementary excitations of materials like Yb$_2$Ti$_2$O$_7$ take the form of extended ``quantum strings" consisting of fluctuating multiple flipped spins connecting monopole pairs$^{25}$.  With inherent quantum dynamics, these novel excitations are extended objects rather than point particles, an interesting feature of quantum spin ice.

To explore the nature of the excitations in quantum spin ice, we performed time-domain terahertz spectroscopy (TDTS) measurements of Yb$_2$Ti$_2$O$_7$ single crystals under magnetic field applied along the $<$001$>$ direction. Spectroscopic features consistent with the picture of quantum strings are observed in the low-field regime. When the field strength increases, a crossover towards field-induced order is observed where the excitations take the form of magnons and two-magnon excitations.

\section{Results}
\subsection{Magneto-optical Measurement}

Magneto-optical measurements are performed in a custom built time domain terahertz (THz) spectrometer. Two single crystals from the same traveling solvent floating zone boule are used in the measurement; they will be referred to as Sample A and B throughout the text. The single crystal Yb$_2$Ti$_2$O$_7$ samples are cut and polished with their surfaces oriented along the  $<$001$>$ plane. The sample is hosted in a 7 T superconducting magnet. Experiments are performed in transmission geometry as illustrated in Fig. 1(b), with temperatures of the samples on the scale of the exchange interaction of Yb$_2$Ti$_2$O$_7$. These temperatures are about one order of magnitude higher than the value of the debated ordering temperature $T_C$, while at the same time low enough that spin correlations are well developed to allow the study of quantum spin ice physics. The incident THz pulse is linearly polarized, with the wavevector \textbf{k} of the pulse defining the \textbf{z} direction. Measurements are performed in both the Faraday geometry (applied magnetic field in the \textbf{z} direction) as well as the Voigt geometry (field in the \textbf{y} direction).

\begin{figure*}
\includegraphics[trim = 10 10 5 10,width=7cm]{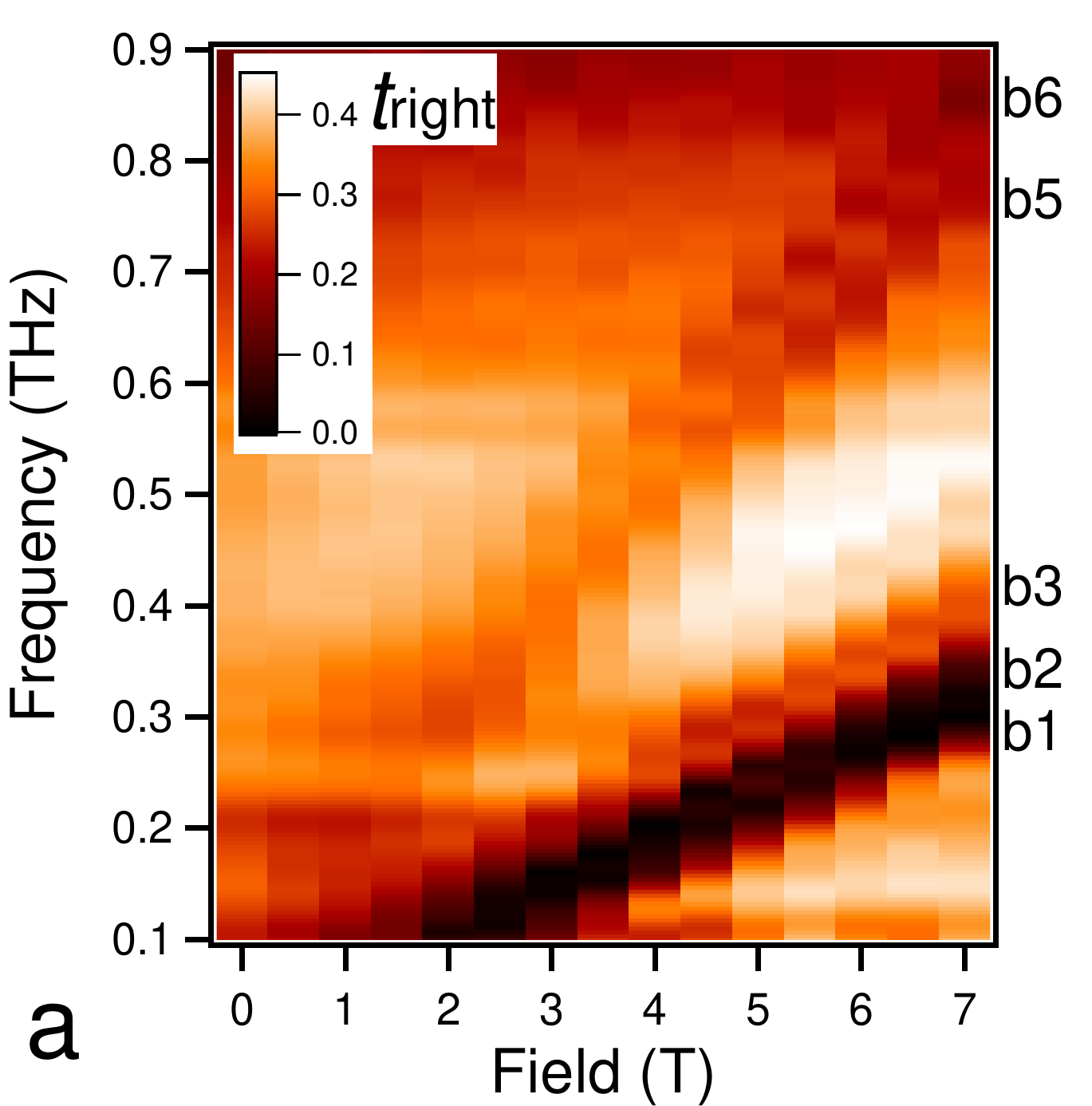}
\includegraphics[trim = 10 10 5 10,width=7cm]{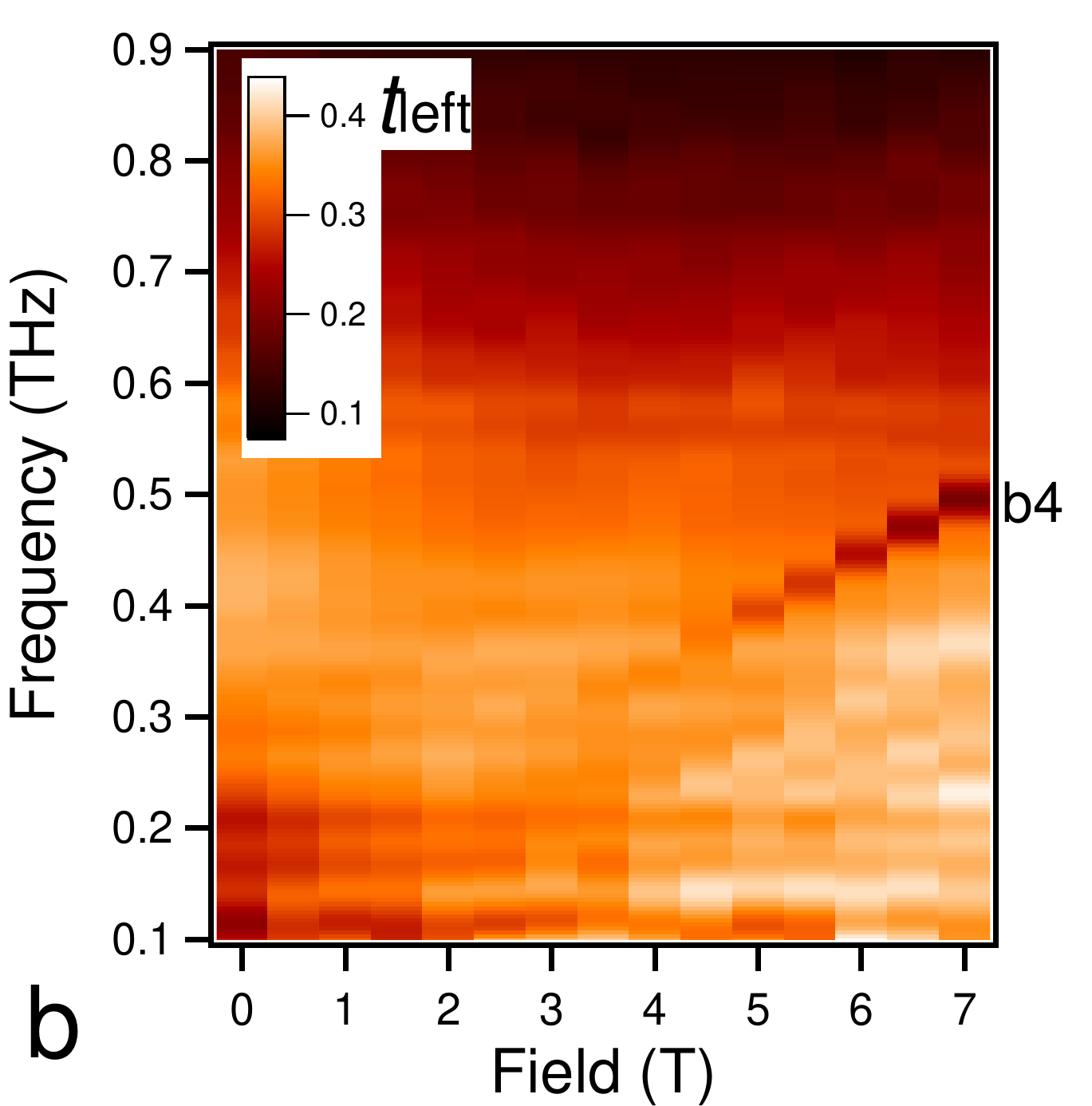}
\label{fig2}
\caption{\textsf{\textbf{Transmission measured in the Faraday geometry}} (a),(b) The intensity plot of the transmission of the (a) right circularly polarized light and (b) left circularly polarized light. Data presented here are measured from Sample B at 4.3 K. Magnetic excitations are marked \textit{b1} to \textit{b6} from low to high frequency. }
\end{figure*}

In magnetic insulators, the complex transmission function $\tilde{t}(\omega)$ we obtain is related to the complex frequency dependent susceptibility $\chi(q=0)$ as $-\ln{t(\omega)} \propto \omega \chi(q=0,\omega)^{26}$.  TDTS measures the $q \rightarrow 0$ response and as such is very useful for materials with interesting $q=0$ correlations$^{27}$. The technique operates at a frequency range most relevant to the physics of quantum spin ice materials (0.4 to 10 meV); it has the strength of high signal to noise, excellent energy resolution, and short acquisition times. Further details can be found in the Methods section.

\subsection{Magnetic Excitations in Faraday Geometry}

In the Faraday geometry, the applied DC magnetic field is oriented along the THz pulse propagation direction, and the incident THz pulse is linearly polarized along the \textbf{x} axis of the laboratory frame. With the spins aligned and precessing around the applied field direction, one naturally expects magnetic absorption to be observed in the right circular polarized (RCP) channel, which is a direct consequence of the fundamental coupling between magnetic field and spin precession. As a result, the magnetic excitations in this geometry would best be studied in the circular polarization frame, which can be measured with a phase sensitive spectroscopic tool such as TDTS$^{28-31}$.

In Fig. 2(a) and (b) we show the transmission magnitude for the right and left circularly polarized (RCP and LCP) light as a function of frequency and field measured at 4.3 K from Sample B in the Faraday geometry (See Methods for transformation to circular base).   The measurements were taken every 0.5 T from 0 to 7 T. Clear dark lines diagonally across the figures immediately show the position of the absorptions in the spectra, revealing the existence of several branches of magnetic excitations. As we show in the supplementary information, spectra taken at 1.65 K and 4.3 K are almost identical, except that the absorptions are slightly sharper and stronger at 1.65 K. Recalling that the quantity we explicitly measure (and are plotting) is related to $\omega \chi(q=0,\omega)$, we observe that these excitations are generally losing spectral weight in $ \chi$ as the field increases and they move to higher $\omega$.  For the RCP channel, careful examination of the data reveals a total of five branches for fields higher than 3 T. They are labeled \textit{b1, b2, b3, b5,} and \textit{b6} from low to high frequency. There is a single LCP branch, \textit{b4}, observed for fields higher than 4 T. This absorption in a \textbf{+z} magnetic field is anomalous and will be discussed in more detail below. While branches \textit{b1} to \textit{b4} have similar slope in the frequency-field plot, suggesting similar effective g-factors; branches \textit{b5} and \textit{b6} clearly have higher g-factors. 



\begin{figure*}
\includegraphics[trim = 10 5 5 5,width=5.5cm]{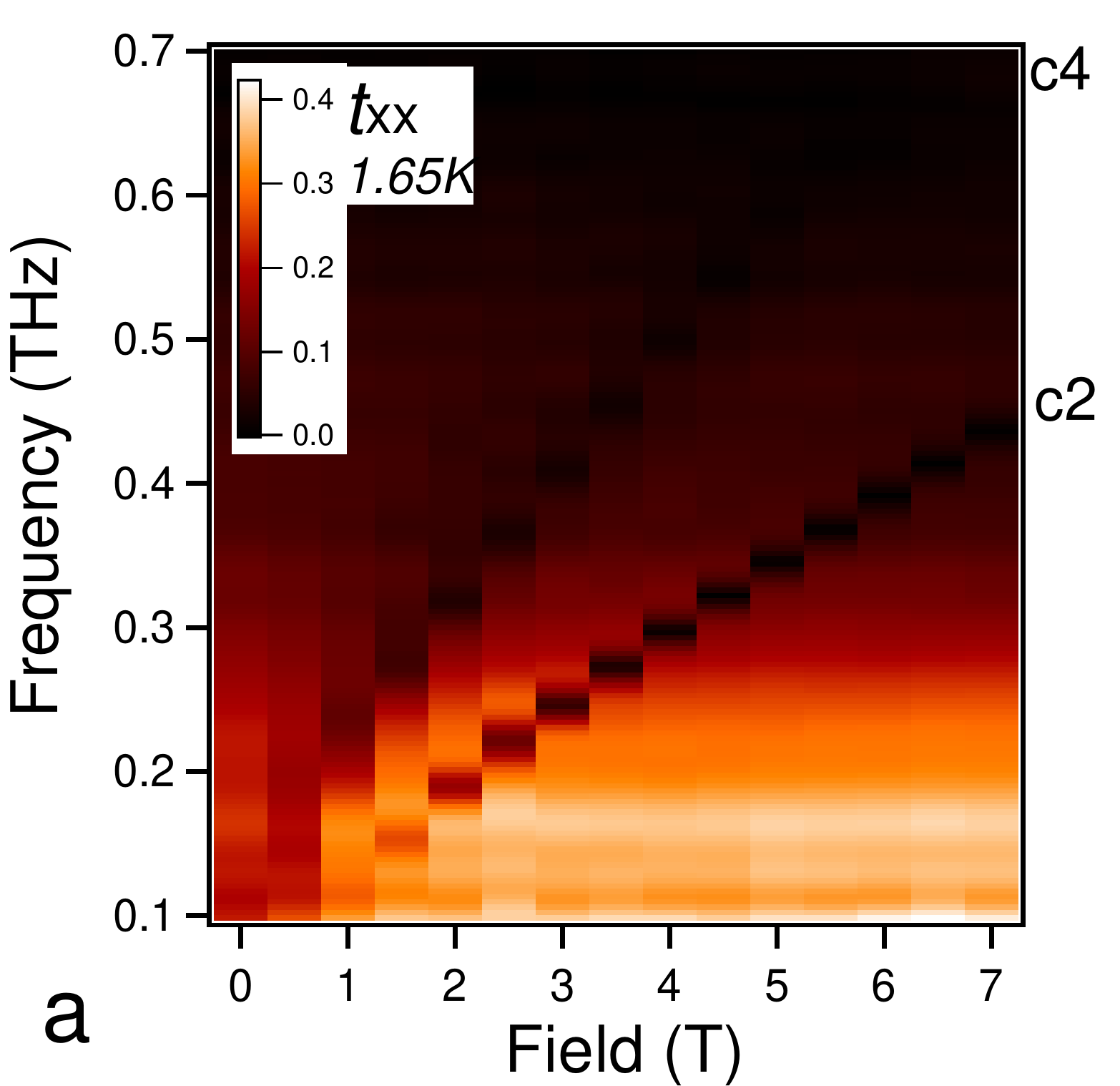}
\includegraphics[trim = 10 5 5 5,width=5.5cm]{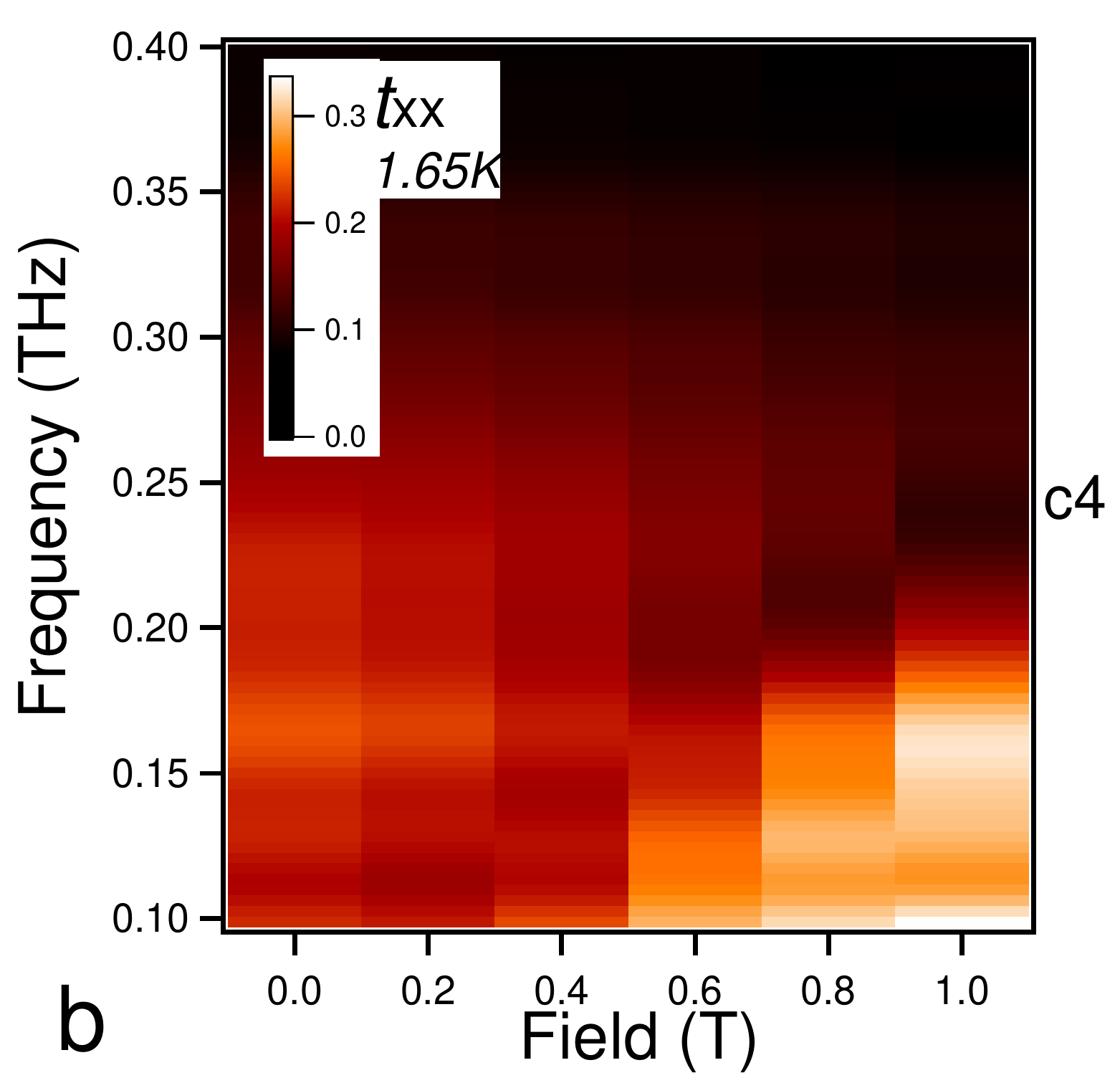}
\includegraphics[trim = 10 5 5 5,width=5.5cm]{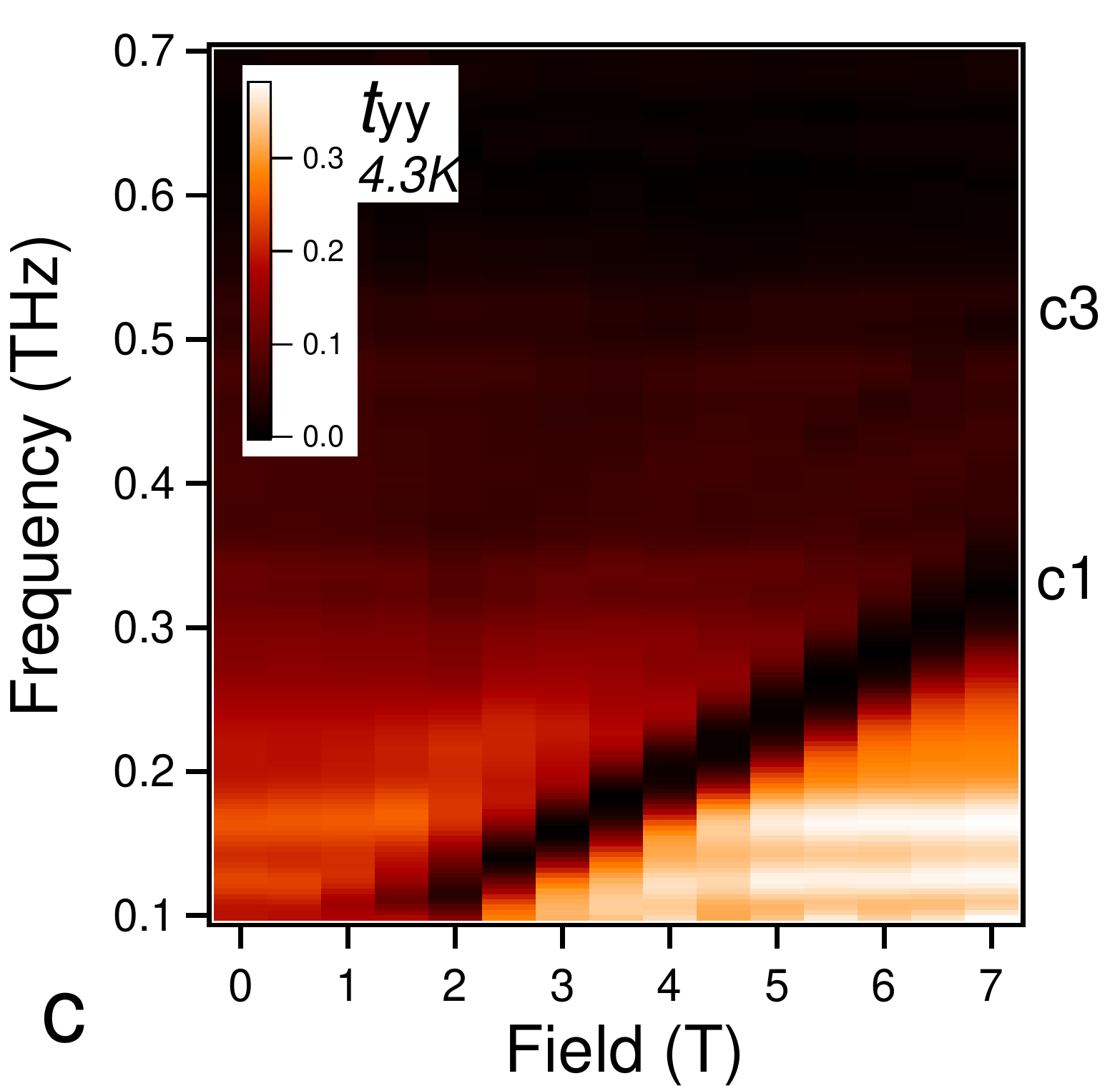}
\label{fig4} 
\caption{ \textsf{\textbf{Transmission as a function of field and frequency in Voigt geometry}} Transmission magnitude as a function of frequency and field measured from Sample A at 1.65 K with the electric polarization of the THz pulse in the \textbf{x} direction ((a), (b)), and \textbf{y} direction measured at 4.3 K (c). For (a) and (c), data were taken from 0 to 7 T with 0.5 T steps, and (b) shows the data below 1 T with 0.2 T steps. Magnetic excitations are labeled \textit{c1} to \textit{c4} from low to high frequency.}
\end{figure*}

In Yb$_2$Ti$_2$O$_7$, there are four magnetic ions in each primitive unit cell; naively one would expect at most four spin waves modes, as has been observed in INS experiments recently$^{14}$. The fact that six different branches of magnetic excitations are observed reveals the unconventional nature of some of these excitations.   Another interesting feature is that as the observed excitations move to low-frequencies at low-field, they become less well defined and eventually turn into a broad magnetic absorption. Similar features are also obtained with sample A, from temperatures 1.65 K upward. All the features other than \textit{b1} and \textit{b2} fade away as temperature increases above 12 K. Details of the temperature dependence as well as additional data can be found in the supplementary information. 

\subsection{Magnetic Excitations in Voigt Geometry}

In the Voigt geometry, the magnetic field is oriented along the \textbf{y} direction. As a result, the circular frame is no longer an appropriate base for the study. Symmetry considerations dictate the absence of circular effects with this geometry$^{31}$.  Shown in Fig. 3(a) and (b) are the transmission magnitudes with electric polarization along the \textbf{x} direction, \textit{$t_{xx}$}, measured in Sample A at 1.65 K, with Fig. 3(b) showing a more detailed measurement below 1 T, while results for \textit{$t_{yy}$} are shown in Fig. 3(c). A total of four branches of magnetic excitations are observed, and labeled \textit{c1} to \textit{c4} from low to high frequency. While \textit{c1} to \textit{c3} show similar g-factors, \textit{c4} shows a higher value. Excitations \textit{c2} and \textit{c4} show a notable downward concavity at low magnetic fields.  As will be discussed in detail below, we believe that these features are consistent with a picture of quantum spin ice where high-field one- and two-magnon states continuously evolve into string excitations at low field.

\section{Discussion}
\subsection{One- and Two-Magnon Features}

As shown in Fig. 2 and 3, the magnetic excitations we observe at fields greater than 3 T in both the Faraday and Voigt geometries can be categorized into two groups: \textit{b1} to \textit{b4} and \textit{c1} to \textit{c3} having a similar, lower value of the g-factor (2.96 to 3.86, values listed in Fig. 4 caption), while \textit{b5}, \textit{b6} and \textit{c4} have larger g-factors (6.30 to 6.74). The frequencies of the observed magnetic excitations (with demagnetization correction applied) are summarized in Fig. 4(a).


\begin{figure*}
\includegraphics[trim = 10 60 0 5,width=18.0cm]{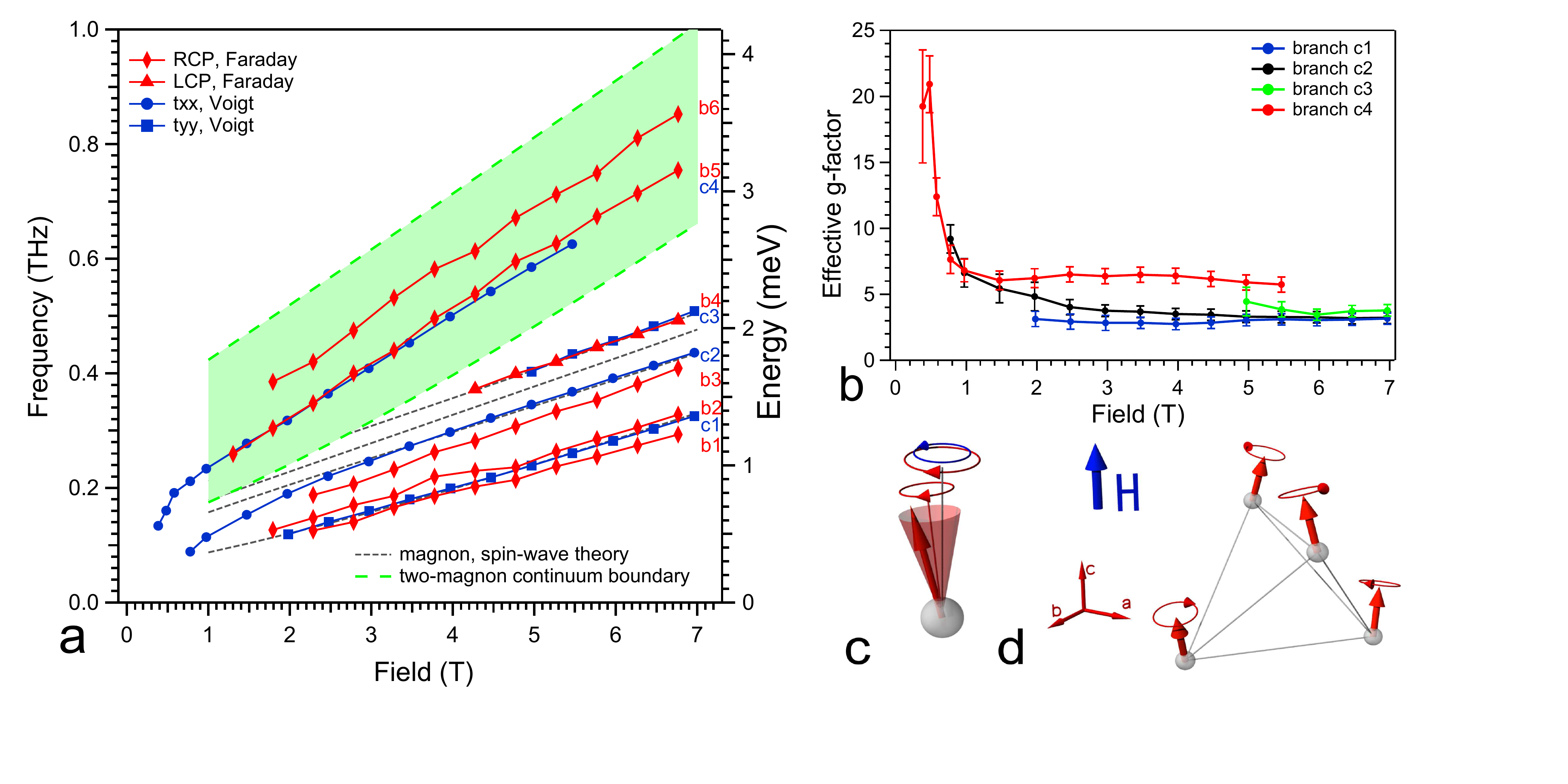}
\label{fig5} 
\caption{ \textsf{\textbf{Summary of the experimental and theoretical results}}  (a) Summary of the magnetic excitation frequencies as a function of field from both Faraday and Voigt geometries. Demagnetization correction has been applied to all the data shown in the figure. Experimental data are shown as solid symbols connected with lines, while dashed lines present results from theoretical calculations. All the data shown in the figure are measured at 1.65 K, except for the \textit{t}$_{yy}$ polarization, which is taken at 4.3 K. Linear fits to data above 3 T yield the effective g-factors of the excitations to be: \textit{b1, b2} and \textit{c1}: 2.96; \textit{b3}: 3.51; \textit{c2}: 3.86; \textit{c3} and \textit{b4}: 3.72; \textit{c4}: 6.30; \textit{b5}: 6.56; \textit{b6}, 6.74. (b) Effective g-factors of branches \textit{c1} to \textit{c4} measured from sample A at 1.65 K. (c) schematic illustration of the spin precession motion with a $<$001$>$ applied field. For a single spin, the equilibrium direction will be slightly tilted away from the field direction. As a result, the precess motion has both a large clockwise component (red arrow) and a small counter-clockwise component (blue arrow) when viewed along the field direction. (d) when the four spins in a single tetrahedron all precess in phase, the resulting magnetic excitation is the observed magnon with the lowest energy with right circular polarization(\textit{b1, b2, c1}).}
\end{figure*}

Upon examination of the complete data set, one notices that the energies of the magnetic excitations with lower g-factors are in the same range as the magnons observed by INS in the field induced ordered state of Yb$_2$Ti$_2$O$_7$ $^{14}$. Although the direction of the applied magnetic field in the INS experiment is different from the present study, this provides a useful hint in establishing the nature of those branches of excitations observed in our TDTS as magnons. 

To form a better understanding of the observed magnetic excitations, we performed a spin-wave analysis with the interaction parameters obtained in Ref. 14 as the starting point.  Linear spin-wave theory predicts a total of four magnon modes, given the four unique spins in the unit cell of Yb$_2$Ti$_2$O$_7$. Only two of the magnon modes are predicted to be visible as $q \rightarrow 0$ in the Faraday geometry, while in the Voigt geometry three modes are predicted to be visible. As shown in Fig. 4, both the energy and the polarization state of the excitations observed in the LCP in the Faraday geometry as well as the ones in the Voigt geometry match the spin-wave theory prediction very well at high fields, where the field induced LRO supports the existence of well defined magnon modes. With the high signal to noise ratio of the TDTS, the data above 3 T from these three branches allowed a further refinement of the exchange parameters in a least squares fitting procedure (see supplementary information for further details).


One surprising aspect of the data was the clear observation of a LCP branch \textit{b4} when the magnetic field was applied in the \textbf{+z} direction (Fig. 2b). As mentioned above, at first sight this seems to be in contradiction to the natural coupling between magnetic field and spin moments.  Previously such LCP spin precession has only been proposed for metallic ferromagnets with strong spin-orbit interaction$^{32}$. This interesting observation finds a natural explanation in the context of the spin-wave theory.  With the easy plane anisotropy of the Yb$^{3+}$ ion, the equilibrium direction of the spin is tilted away from the $<$001$>$ axis, moreover, the spin precession is slightly elliptical around the applied field in the $<$001$>$ direction.   Thus, the precession motion has both LCP and RCP components when viewed along the field direction. This is schematically illustrated in Fig. 4(c) and (d). For the highest-energy magnon mode, which corresponds to the \textit{b4}/\textit{c3} branches, the phase relationship between the spins in the unit cell is such that the RCP components cancel exactly, leaving only the LCP component in the Faraday geometry. 

Another interesting feature of the data is the observation of excitations with clearly higher g-factors than the magnons. A closer examination of the data reveals that excitations \textit{b5}, \textit{b6} and \textit{c4} have a g-factor about twice the value of the magnon branches.  Interestingly they also appear in the middle of the two-magnon continuum band obtained from the spin-wave calculation, as shown in Fig. 4(a). These suggest that, while the branches \textit{b1} to \textit{b4} and \textit{c1} to \textit{c3} can be understood as magnons, \textit{b5}, \textit{b6} and \textit{c4} are two-magnon excitations.  The fact that they appear in the middle of the calculated two-magnon continuum suggests they are unlikely to be simple two-magnon bound states. Thus it remains to be understood how these excitations can be stable and have well defined energies.  Moreover our spin-wave calculations show that these excitations do not match any peaks in the joint two-magnon density of states.  Note that the two-magnon continuum band intersects with the single magnon modes at low fields. This gives the possibility of decay for the higher energy single magnons at low fields.

Also at odds with the spin-wave calculation is the observation of four single magnon-like branches in the Faraday geometry, as only the highest and the lowest modes are expected to be visible. In comparing the geometries, \textit{b1}, \textit{b2}, and \textit{c1} all appear in the energy range of the calculated lowest magnon mode. Our data suggest that  \textit{b1} and \textit{b2} are distinct features (more clearly shown in Fig. SI1 and SI5), although they are difficult to distinguish due to the strong absorption in this frequency range.  They may derive from the same lowest-energy magnon state, but with some unexplained splitting.  Moreover, it is difficult to understand \textit{b3} as the second or third highest magnons that have picked up a finite intensity by some means (from misalignment of the applied field for instance). Its energy is below the \textit{c2} excitation in the Voigt geometry, which is straightforwardly assigned to the second highest magnon mode. These discrepancies which appear only in the Faraday geometry are as of yet unexplained. Further details of the spin-wave analysis of the four magnon modes are given in the supplementary information.

\subsection{The Crossover to Quantum Strings}

As field decreases, the match between the spin-wave calculation and the experiments becomes progressively worse (Fig. 4(a)). At the temperature range of the study, spin correlations are well developed and at low fields the material approaches the nominal quantum spin ice regime$^{12, 33}$. Thus it is not surprising to see the spin-wave theory fails at low fields.  The spectroscopic features in this region reveal the nature of the excitations unique to the quantum spin ice state.  Among those features, the most prominent one is the change of slope of the magnetic excitations at low fields.  We believe this large concave downward non-linearity is a general signature of quantum mixing of states with different numbers of spin flips. When the field is lowered from the high-field regime (where the excitations should be understood as spin flips of a quantized number e.g. magnons) the excitations of the system are now expected to be admixtures of chains of flipped spins with different lengths, as proposed in the quantum string picture and also observed in spin chains $^{34}$. In this scenario one expects that as the field diminishes, strings get longer. As a result their g-factor increases and the slope of the curve goes up. The strong enhancement of these g-factors below 2 T clearly observed for branches \textit{c2} and \textit{c4} as field decreases provides strong support for this picture, as shown in Fig. 4(b). This strong increase of g-factors is evidence of significant admixing of states of many flipped spins in this part of the phase diagram.

The quantum string picture was derived in the low-field limit and requires the breaking of time-reversal symmetry either by a $<$001$>$  directed field or a spontaneous magnetization in this direction$^{25}$. It was 
 proposed that a hierarchy of states would be observed corresponding to strings of different lengths. However, in real materials with multiple spins in the unit cell and finite non-Ising exchange terms, such a hierarchy of states might be difficult to realize explicitly due to the decay channels opening up with overlapping bands of multi-string and multi-magnon continuums (as implied in Fig. 4(a)). In the present study high-field two-magnon excitations are observed to evolve continuously into string states at low fields, although their stability remains to be understood. In any case, one may still expect that the lowest magnon features observed at high fields evolve into string-like states at low fields.  Our observations are consistent with such a picture that the lowest elementary excitations of a quantum spin ice in small magnetic fields are quantum strings connecting weakly bound monopoles.  As discussed above, the application of a symmetry breaking field can stabilize excitations whose properties reflect certain aspects of an anomalous state without long-range order.  In the present case, the observation of a crossover from magnons to strings as the field is decreased reflects the presence of deconfined monopoles in the zero-field quantum spin ice state.

\section{Methods}
\subsection{Sample Fabrication and Characterization}
Single crystal samples of Yb$_2$Ti$_2$O$_7$ were grown at McMaster University with the optical floating zone technique. Details of the sample growth are discussed elsewhere$^{17}$. Samples A and B used for the THz measurements are two single crystals with their largest surface oriented parallel to the  $<$001$>$ plane. Sample A has a circular cross section with a diameter of 4.2 mm, with thickness of 0.725 mm, while Sample B has a rectangular cross section that measures 5.46 by 2.37 mm, with a thickness of 0.648 mm. Both samples are transparent. Laue x-ray diffraction was performed to assure the samples used do not contain domains of misaligned microstructure or crystallites.

The low temperature heat capacity of Sample B has been measured from 150 mK to 800 mK, and shows the same behavior as the non-annealed crystal reported in Ref. 17. As shown in Ref. 33 and 35, at temperatures of our current study, the exact nature of the low temperature state is not relevant, as the entire phase diagram is expected to be in the \textit{thermal spin liquid} state at temperatures on the scale of $J_{ZZ}$. Also, as reported in Ref. 17, 20, and 21, sample variations are minimum for temperatures above 1 K, thus the possible sample variation does not affect our conclusion in this paper. 

\subsection{Time Domain Terahertz Spectrometer Set Up}
A home-built time-domain terahertz spectroscopy was set up at Johns Hopkins University with two dipole switches as THz emitter and receiver. The sample was hosted in a cryogen free split coil superconducting magnet. The magnet provides magnetic fields up to 7 T, and a base temperature down to 1.6 K for the measurements.  The magnet has four windows allowing for optical access, and can be switched between Faraday and Voigt geometries. The single-crystal samples were mounted on a metal aperture. The electric field profiles of the terahertz pulses transmitted through the sample and an identical bare aperture were recorded as a function of delay stage position, and then converted to time traces. FFTs of the sample and aperture scans gave the complex transmitted and incident pulse spectra, from which the complex transmission coefficients were obtained. 

\subsection{Polarization Modulation Technique in Terahertz Polarimeter}
As shown in the previous sections, the simultaneous measurement of two orthogonal components of the transmitted THz pulse is critical for this experiment. This is realized by a terahertz polarimeter with polarization modulation technique based on a rotating polarizer developed recently$^{28}$. The main set up contains three polarizers, one rotates at angular velocity $\Omega$ that modulates the polarization state of the THz pulse to be measured (shown in Fig. 1(b)); another one downstream oriented with the transmission axis parallel to the \textbf{x} axis. Another polarizer upstream creates a linearly polarized incident pulse before the sample.

The effect of the two polarizers in the polarimeter on an arbitrary electric field vector can be expressed as a product of two matrices:

\begin{equation}
E_{final}=P_X P_{\Omega t} \binom{\tilde{E}_x}{\tilde{E}_y}
\end{equation}
where:
\begin{equation}
P_X=\begin{pmatrix}
 1& 0\\ 
 0& 0
\end{pmatrix};P_{\Omega t}=\begin{pmatrix}
\cos^{2}{\Theta} & \cos{\Theta}\sin{\Theta}\\ 
\cos{\Theta}\sin{\Theta} & \cos^{2}{\Theta}
\end{pmatrix}
\end{equation}
straightforward algebra gives:
\begin{equation}
\boldsymbol{E_{final}}=\frac{1}{2}\binom{\tilde{E}_x(1+\cos{2\Theta})+\tilde{E}_y \sin{2\Theta}}{0}
\end{equation}

Here $\Theta$=$\Omega$t+$\phi_0$ is the angle of the rotating polarizer with respect to the \textbf{x} axis, $\phi_0$ is the initial angle. The signal detected after the polarimeter is then fed into a lock-in amplifier with a reference signal of frequency 2$\Omega$. In this way, the in-phase and out-of-phase readings of the lock-in amplifier measure $\tilde{E}_x$ and $\tilde{E}_y$ simultaneously, with an appropriate choice of $\phi_0$. Note that it is important that the sense of rotation (direction of the angular velocity \textbf{$\Omega$}) of the rotating polarizer be the same as the THz pulse propagation direction; otherwise instead of $\tilde{E}_y$, $-\tilde{E}_y$ will be measured, which would give a reversed definition of right/left circular polarization. See Ref. 25 for further details on this technique.

\subsection{Transforming into Circular Base in the Faraday Geometry}

Following the previous section, when $\tilde{E}_x$ and $\tilde{E}_y$ of the transmitted pulse are measured while retaining their phases, it is possible to transform the frame of reference into circular polarized frame, thus providing an opportunity to study the magnetic excitations in their natural circularly polarized basis in the Faraday geometry. With $\tilde{E}_r$, $\tilde{E}_l$ standing for the electric field component in RCP and LCP channels, we have:
\begin{equation}
\tilde{E}_r = \frac{1}{\sqrt{2}}(\tilde{E}_x+i\tilde{E}_y); \\
\tilde{E}_l = \frac{1}{\sqrt{2}}(\tilde{E}_x-i\tilde{E}_y)
\end{equation}
More details of this transformation of bases can be found in the supplementary information.




\section{addendum}
 This work at JHU was supported by the Gordon and Betty Moore Foundation through Grant GBMF2628 to NPA and the DOE through DE-FG02-08ER46544.  The crystal growth work at McMaster was supported by NSERC.  We would like to thank C. Broholm and J. Deisenhofer for helpful conversations.

\textit{Competing Interests: }The authors declare that they have no competing financial interests.

\textit{Correspondence: }Correspondence and requests for materials
should be addressed to NPA (email:npa@pha.jhu.edu).

\section{Author Contribution}

 LP performed the THz experiments and analysis with assistance from CMM.  SKK, AG, and OT performed the theoretical calculations.   KAR, EK, BDG, provided the high quality single crystals.  SMK cut and aligned the crystals.  NPA conceived and directed the project.  All authors contributed to discussions on data analysis and writing of the manuscript.

\newpage.

\section{Supplementary Information}
\subsection{Determination of Transmission Matrix in Linear Base}

Here we provide the detailed Jones matrix analysis of the time domain terahertz spectroscopy measurement, and the analysis of the transmission matrix in the linear and circular bases.  With the direction of the THz pulse propagation (wave vector \textbf{k} direction) set as the \textbf{z} axis, the electric field vector at any point along the optical path can be written as a column vector with complex entries.

\begin{equation*}
\mathbf{\tilde{E}}=\binom{\tilde{E}_x}{\tilde{E}_y}  
\end{equation*}

The transmission of the THz pulse through the sample can be viewed as a transformation of the incident vector by a 2 by 2 complex transmission matrix:
\begin{equation*}
\binom{\tilde{E}_{x,out}}{\tilde{E}_{y,out}}=\begin{pmatrix}
 \tilde{t}_{xx} & \tilde{t}_{yx} \\ 
 \tilde{t}_{xy} & \tilde{t}_{yy} 
\end{pmatrix}\binom{\tilde{E}_{x,in}}{\tilde{E}_{y,in}}
\end{equation*}

In the Faraday geometry, since the sample is oriented with surface parallel to the  $<$001$>$ plane, the transmission matrix should preserve the 4-fold rotational symmetry of the cubic crystal along the surface normal. Requiring the transmission matrix to remain unchanged after a rotation of 90$^\circ$ along \textbf{z} direction shows that the transmission matrix must be antisymmetric, that is: $\tilde{t}_{xx}=\tilde{t}_{yy}$, and $\tilde{t}_{xy}=-\tilde{t}_{yx}$. Now the transmission matrix in the linear laboratory frame (frame \textbf{xyz}) has only two independent component, and can be determined by:
\begin{equation}
\tilde{t}_{xx}=\frac{\tilde{E}_{x,out}\tilde{E}_{x,in}+\tilde{E}_{y,out}\tilde{E}_{y,in}}{\tilde{E}_{x,in}^{2}+\tilde{E}_{y,in}^{2}}
\end{equation}
\begin{equation}
\tilde{t}_{xy}=\frac{\tilde{E}_{y,out}\tilde{E}_{x,in}-\tilde{E}_{x,out}\tilde{E}_{y,in}}{\tilde{E}_{x,in}^{2}+\tilde{E}_{y,in}^{2}}
\end{equation}
With the \textbf{x} axis set in the direction of the polarization of the incident THz pulse, $\tilde{E}_{y,in}$ would be identically zero.

In a real time-domain terahertz spectrometer, a weak non-zero $\tilde{E}_{y,in}$ can be present due to small imperfections in alignment that give ellipticities in the incoming beam.  For instance they may come from the modification of the incident pulse polarization due to the optical components along the beam path. With the full Jones matrix analysis presented above, a clean separation and calculation of $\tilde{t}_{xx}$ and $\tilde{t}_{xy}$ can be obtained. 

\subsection{Determination of Transmission Matrix in Circular Base}

As discussed in the main text, when magnetic excitations are present in the Faraday geometry, circular base is a more natural choice to analyze the data. Following the discussion in Refs. 26, 28 we present in detail the determination of transmission matrix in the circular base.

\begin{figure}[t]

\includegraphics[trim = 10 0 10 10,width=7cm]{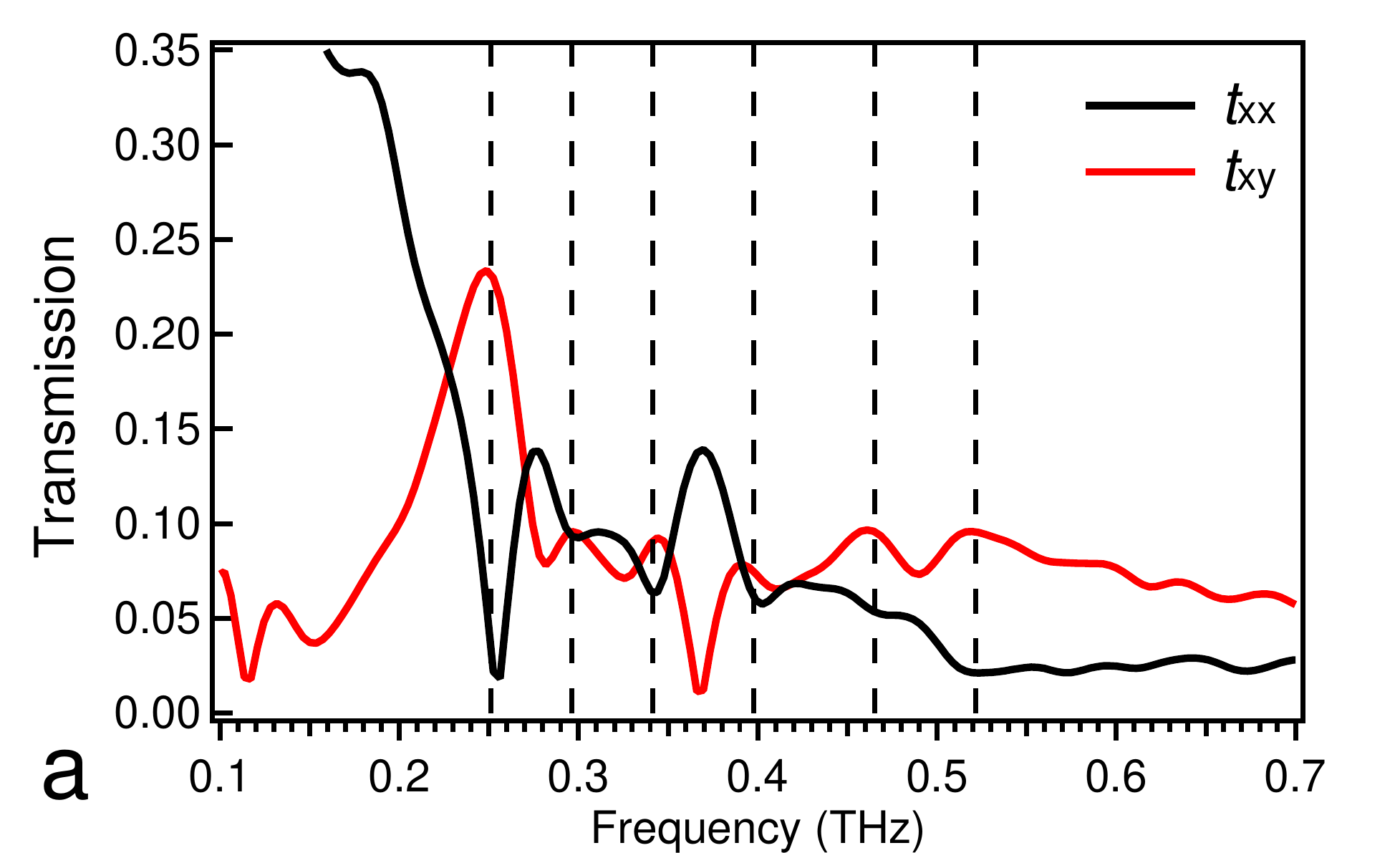}
\includegraphics[trim = 10 0 10 10,width=7cm]{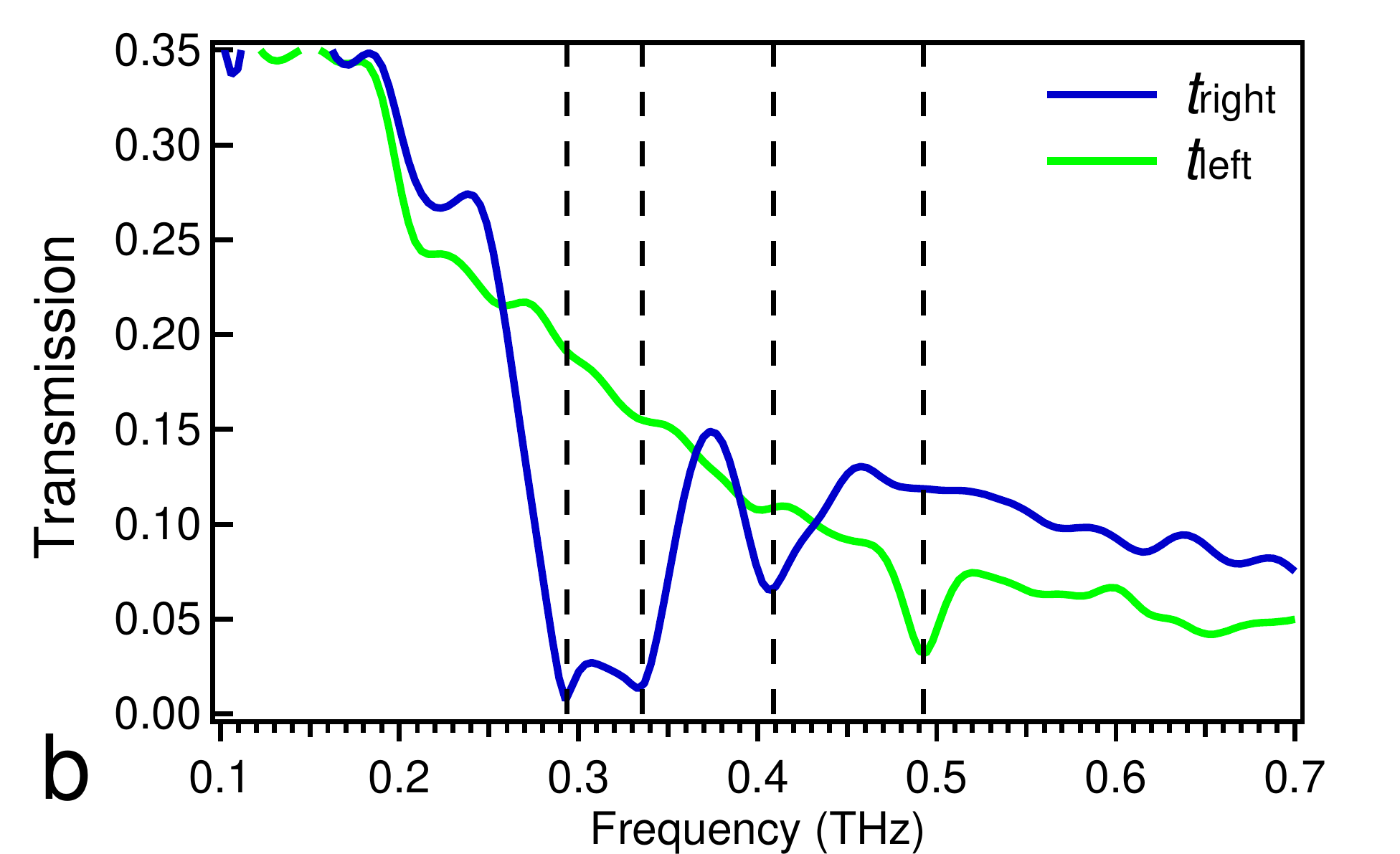}
\label{figSI1}
\setcounter{figure}{0}
\caption{\textsf{\textbf{Change of basis from linear to circular eigenstates}} (a),(b) Typical transmission spectra decomposed in (a) linear base and (b) circular base, measured from Sample A at 1.65 K with applied field of 7 T. Vertical dashed lines denote spectroscopic features.}
\end{figure}

Suppose in the linear base, the electric field vectors are written as \textbf{E$_{in}$} and \textbf{E$_{out}$} for the incident and transmitted pulses; while in the circular base we have \textbf{E$_{in,cir}$} and \textbf{E$_{out,cir}$}. If we write the transformation matrix from the circular to linear base as $\Delta$, then we have:
\begin{equation*}
\mathbf{E_{in}}=\Delta \mathbf{E_{in,cir}}; \mathbf{E_{out}}=\Delta \mathbf{E_{out,cir}}
\end{equation*}
in the linear base with the transmission matrix \textbf{T}, we have
\begin{equation*}
\mathbf{E_{out}}=\mathbf{T} \mathbf{E_{in}}
\end{equation*}
then, follows from those equations, we have
\begin{equation}
\mathbf{E_{out,cir}}=\Delta^{-1} \mathbf{T}\Delta\mathbf{E_{in,cir}}
\end{equation}
clearly now \textbf{$\Delta^{-1}$}\textbf{T}\textbf{$\Delta$} is the new transmission matrix for the circular base.

With $\tilde{E}_r$, $\tilde{E}_l$ denoting the electric field component in the right/left circular polarization channel, we have:
\begin{equation}
\tilde{E}_r = \frac{1}{\sqrt{2}}(\tilde{E}_x+i\tilde{E}_y); \\
\tilde{E}_l = \frac{1}{\sqrt{2}}(\tilde{E}_x-i\tilde{E}_y)
\end{equation}
combined with:
\begin{equation*}
\binom{\tilde{E}_x}{\tilde{E}_y}=\mathbf{\Delta}\binom{\tilde{E}_r}{\tilde{E}_l}
\end{equation*}
gives the transformation matrix:
\begin{equation}
\mathbf{\Delta}=\frac{1}{\sqrt{2}}\begin{pmatrix}
1 & 1\\ 
-i & i
\end{pmatrix}
\end{equation}
and the transmission matrix in the circular frame:
\begin{equation}
T_{cir}=\Delta^{-1}\mathbf{T}\Delta = \begin{pmatrix}
\tilde{t}_{xx}+i\cdot\tilde{t}_{xy} & 0\\ 
 0 & \tilde{t}_{xx}-i\cdot\tilde{t}_{xy}.
\end{pmatrix}
\end{equation}
This matrix is diagonal in the bases of right/left circular polarized light. The diagonal elements gives $\tilde{t}_{right}$ and $\tilde{t}_{left}$, which can be calculated once $\tilde{t}_{xx}$ and $\tilde{t}_{xy}$ are known. Another way of obtaining $\tilde{t}_{right}$ and $\tilde{t}_{left}$ is by decomposing the incident and transmitted THz pulse into right/left circular bases with the transformation matrix \textbf{$\Delta^{-1}$}, then taking the complex ratio of the corresponding spectra. Those two methods are equivalent.

In the Faraday geometry, the magnetic field is oriented in the same direction as the \textbf{k} propagation direction of the THz pulse (\textbf{z} direction in Fig. 1(b)). With the polarization modulation technique, the two orthogonal components (along \textbf{x} and \textbf{y} axes) of the transmitted THz pulse can be obtained $simultaneously$ in a single measurement$^{25}$. This allows the simultaneous determination of two complex transmission coefficients \textit{$\tilde{t}_{xx}$} and \textit{$\tilde{t}_{xy}$}. As mentioned above, for the Faraday geometry, a 4-fold rotational symmetry is expected along the surface normal direction with the FCC symmetry of the material, which requires the transmission matrix to have only two independent components:

\begin{eqnarray}
\tilde{t}_{linear}=\begin{pmatrix}
\tilde{t}_{xx} & \tilde{t}_{xy}\\
-\tilde{t}_{xy} & \tilde{t}_{xx}
\end{pmatrix}
\end{eqnarray}

\noindent Thus the measurement of \textit{$\tilde{t}_{xx}$} and \textit{$\tilde{t}_{xy}$} completely determines the transmission matrix in the linear basis.



Following the discussion above, the transmission matrix can be diagonalized in the circular basis:
\begin{eqnarray}
\tilde{t}_{circular}=\begin{pmatrix}
\tilde{t}_{right} & 0\\
0 & \tilde{t}_{left}
\end{pmatrix}
\end{eqnarray}
where \textit{$\tilde{t}_{right}$} and \textit{$\tilde{t}_{left}$} are the complex transmission coefficient of the right/left circular polarized light. 
 
An example of the transformation of basis is shown in Fig. SI1(a) and (b). In Fig. SI1(a) we plot the magnitude \textit{$|t_{xx}|$} and \textit{$|t_{xy}|$} from 0.1 to 0.7 THz measured from Sample A at 1.65 K with an applied field of 7 T.  Fig. SI1(b) shows the same set of data in the circular base. Dashed lines in the figure mark the position of spectroscopic features. It is immediately evident that the circular base presents the data in a more straightforward fashion.  It is clear that the pairs of dips and peaks in \textit{$t_{xx}$} and \textit{$t_{xy}$} corresponds to absorptions in the right/left circular channel, seen as the dips in \textit{$t_{right}$} and \textit{$t_{left}$}, respectively. The fact that those absorptions show up in the spectra of one type of circular polarized light only clearly demonstrates their nature as magnetic excitations. Another important point is that for a spectra as rich as the ones shown here, analyzing the data in an incorrect frame might lead to the misidentification of excitations. We should emphasize that the use of the time domain technique is vital for the types of analysis reported in this paper, as it allows the determination of both amplitude as well as phase of the terahertz pulses. The transformation between linear and circular bases is possible only when the phase information is retained.

From the above discussion, it is evident that by analyzing the transmission in the circular bases, we can separate contributions from different channels in the magnetic transitions. This provides a huge advantage when trying to understand the nature of those excitations, as we demonstrated in the main text of this paper. Again, measurement with a phase insensitive technique does not allow such a transformation of bases.  

\subsection{Temperature dependence of the spectra in the Faraday geometry}

\begin{figure*}
\includegraphics[trim = 10 5 5 5,width=4.2cm]{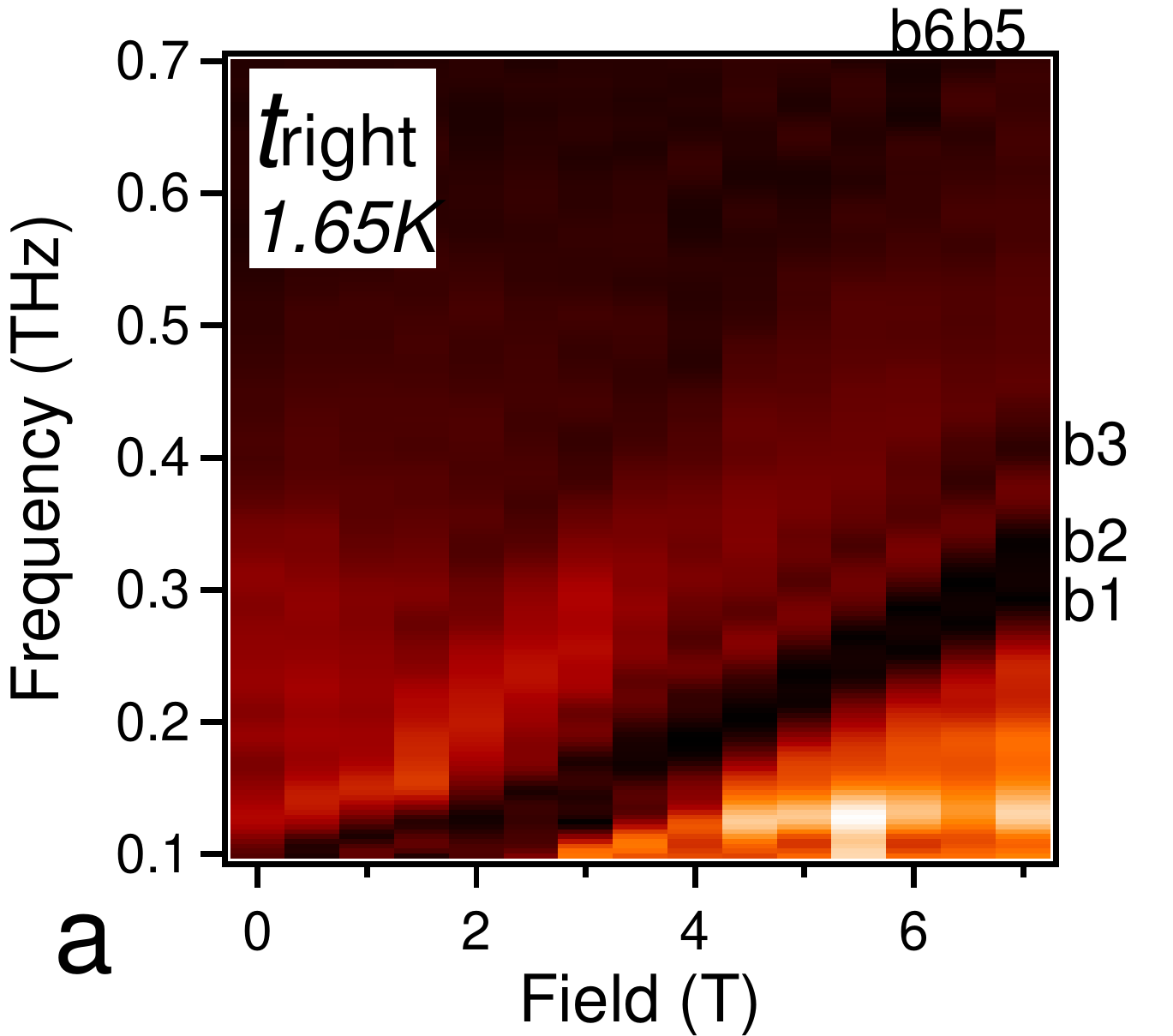}
\includegraphics[trim = 10 5 5 5,width=4.2cm]{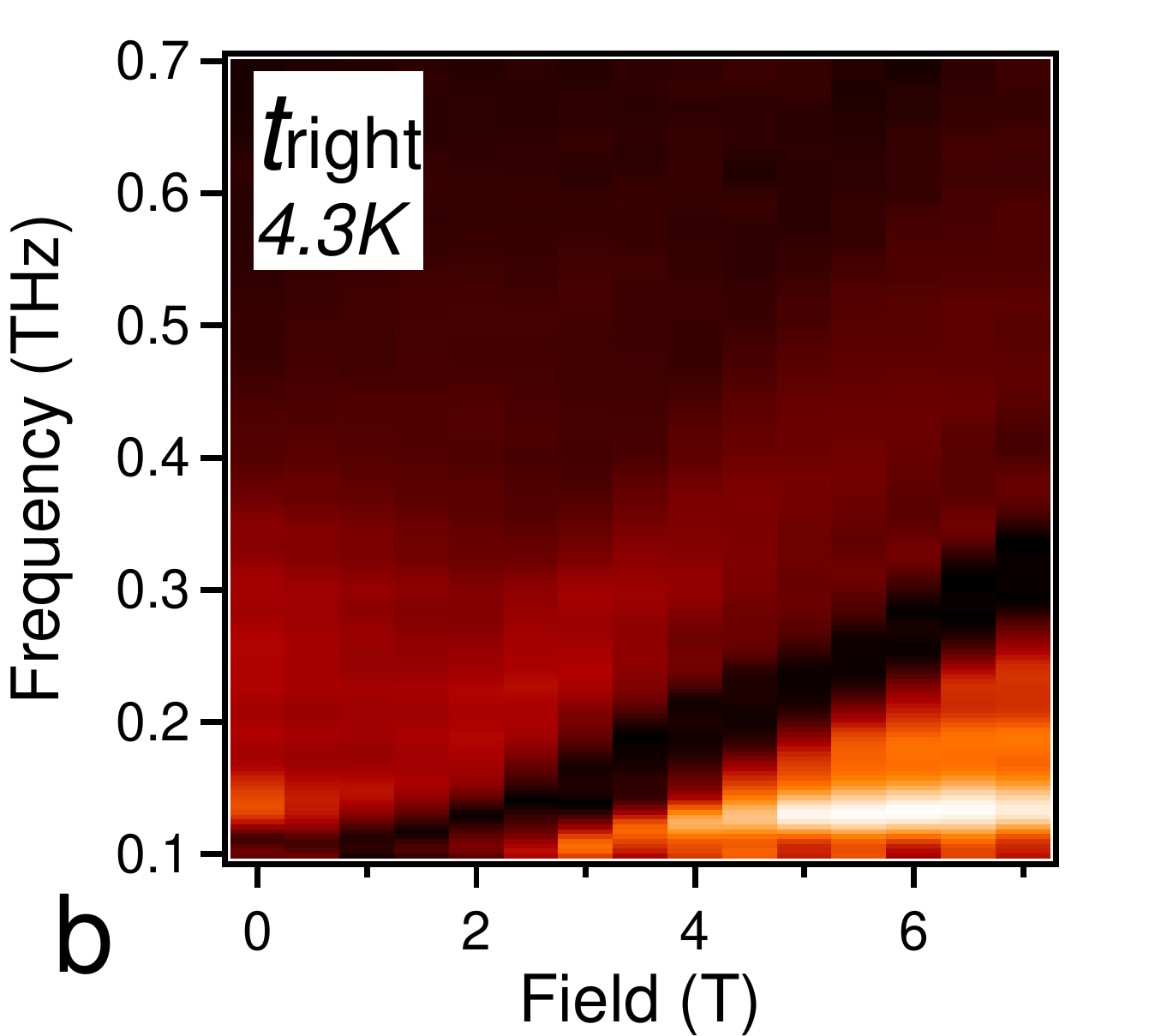}
\includegraphics[trim = 10 5 5 5,width=4.2cm]{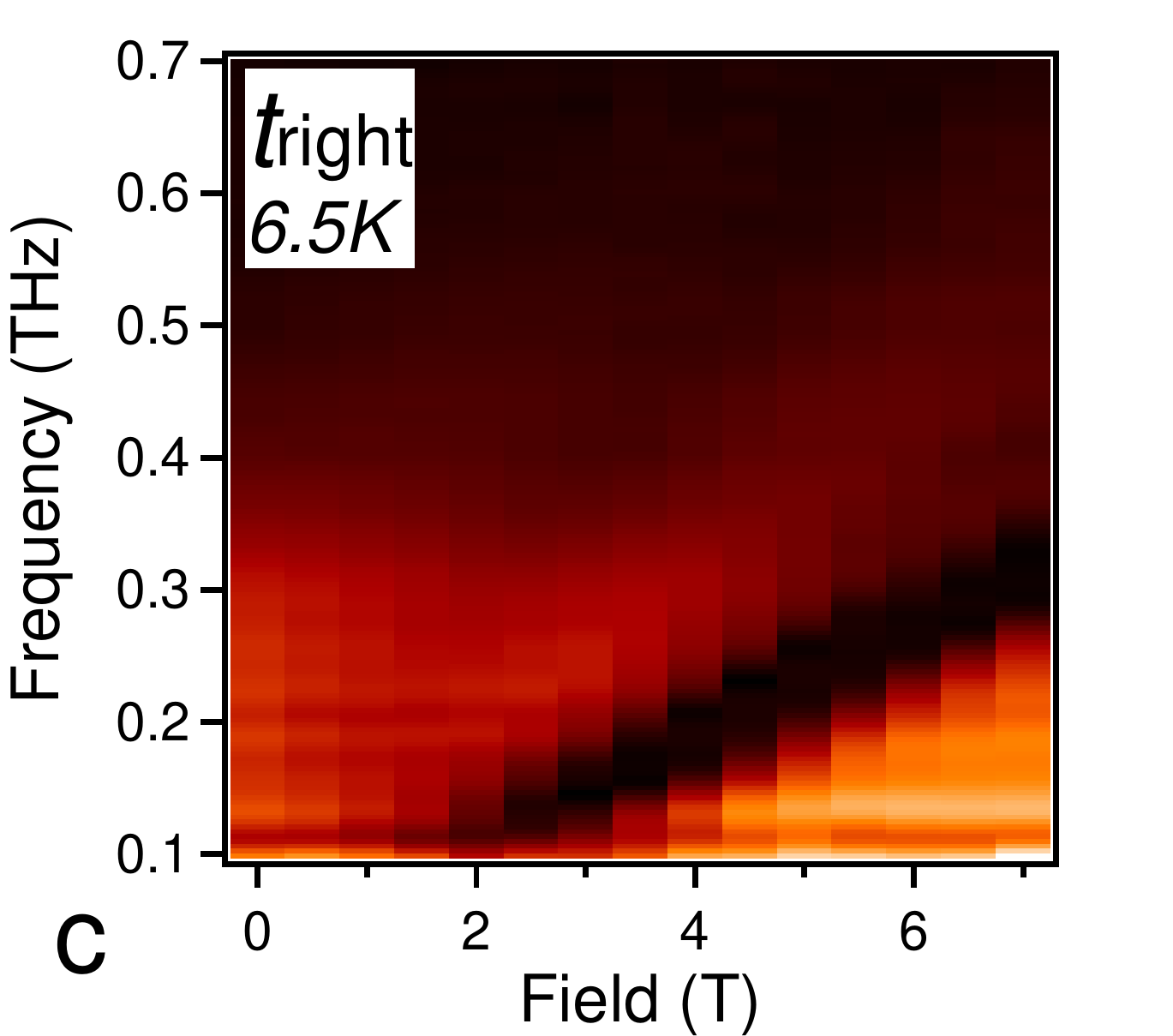}
\includegraphics[trim = 10 5 5 5,width=4.2cm]{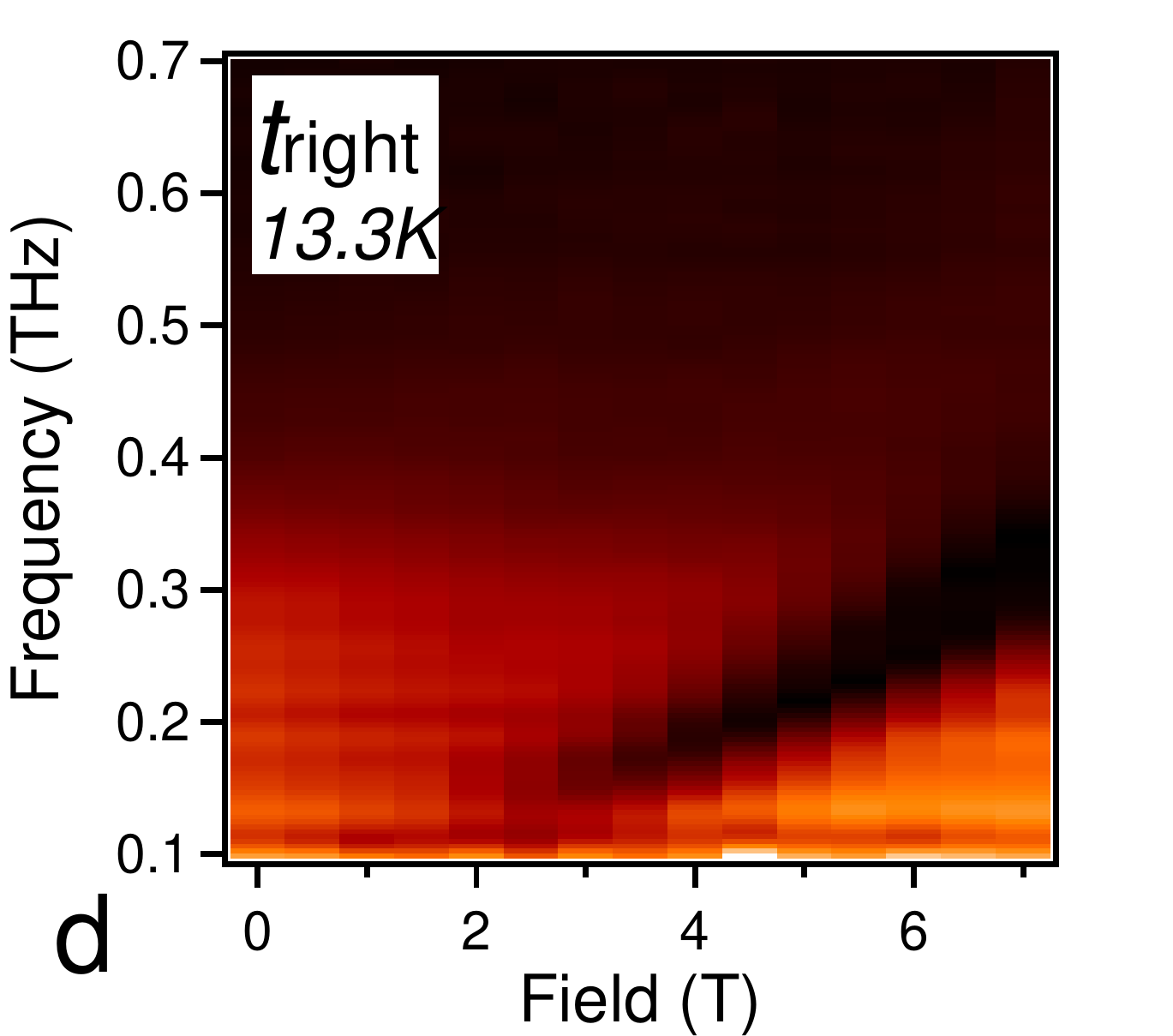}
\includegraphics[trim = 10 5 5 5,width=4.2cm]{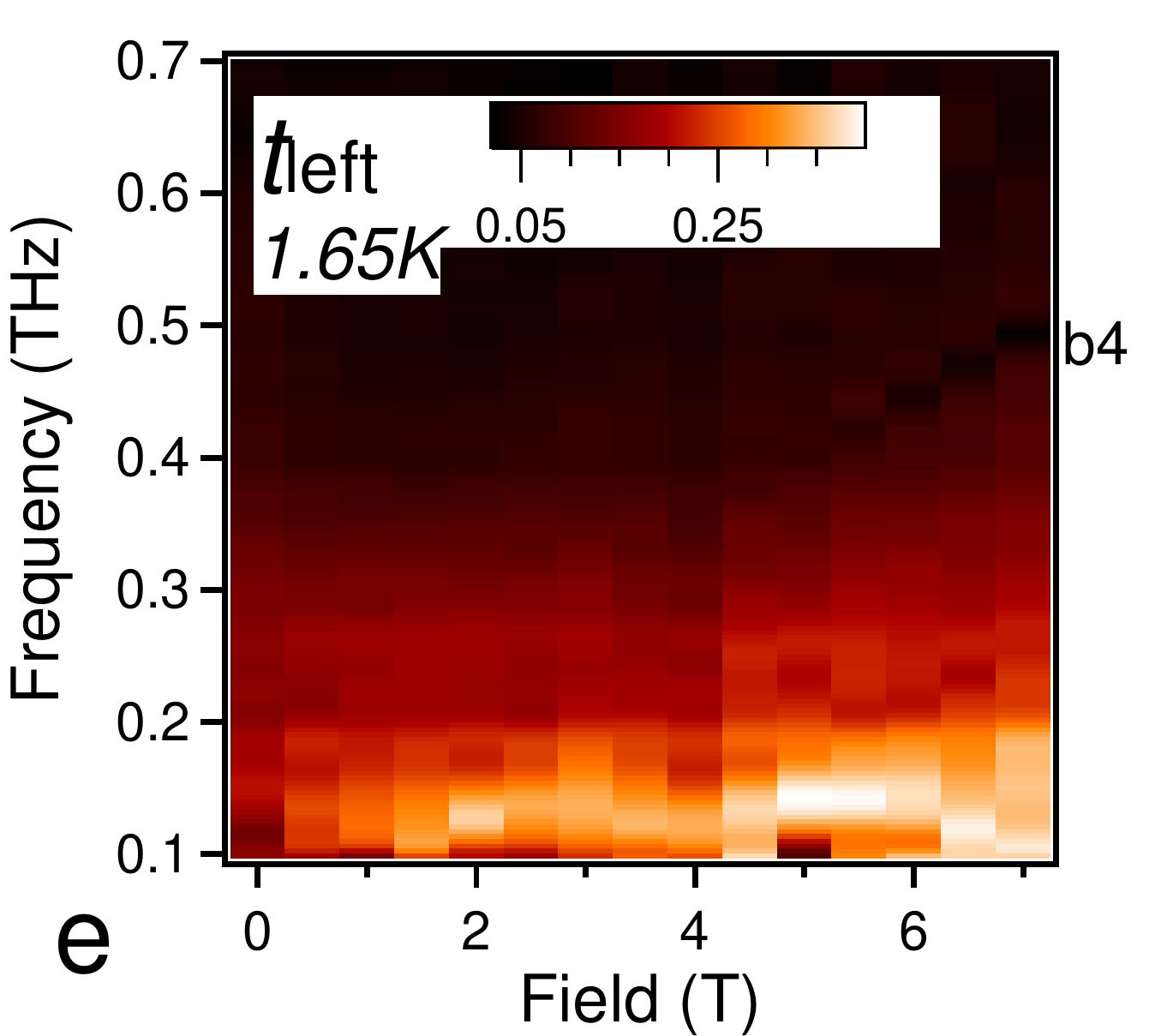}
\includegraphics[trim = 10 5 5 5,width=4.2cm]{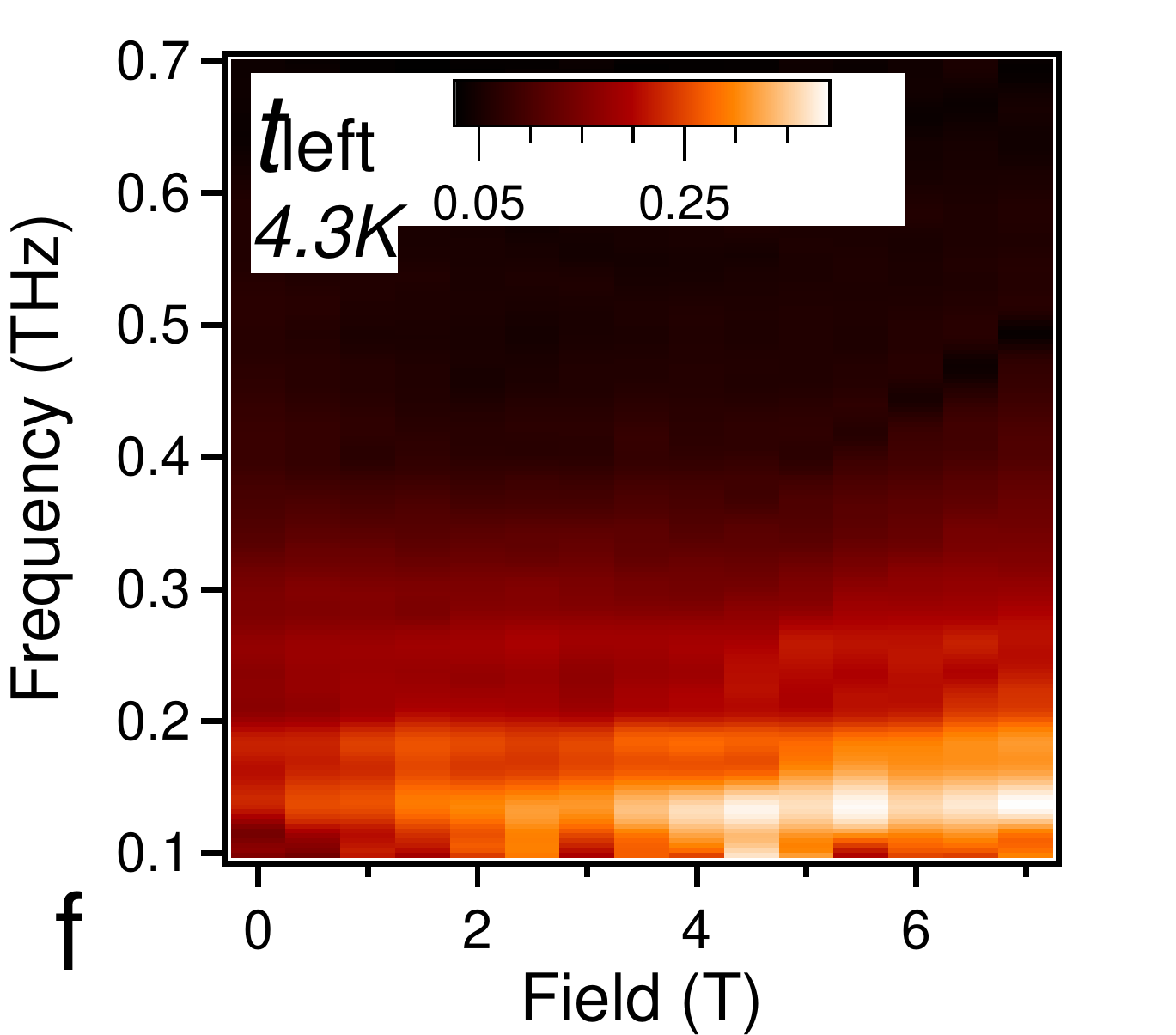}
\includegraphics[trim = 10 5 5 5,width=4.2cm]{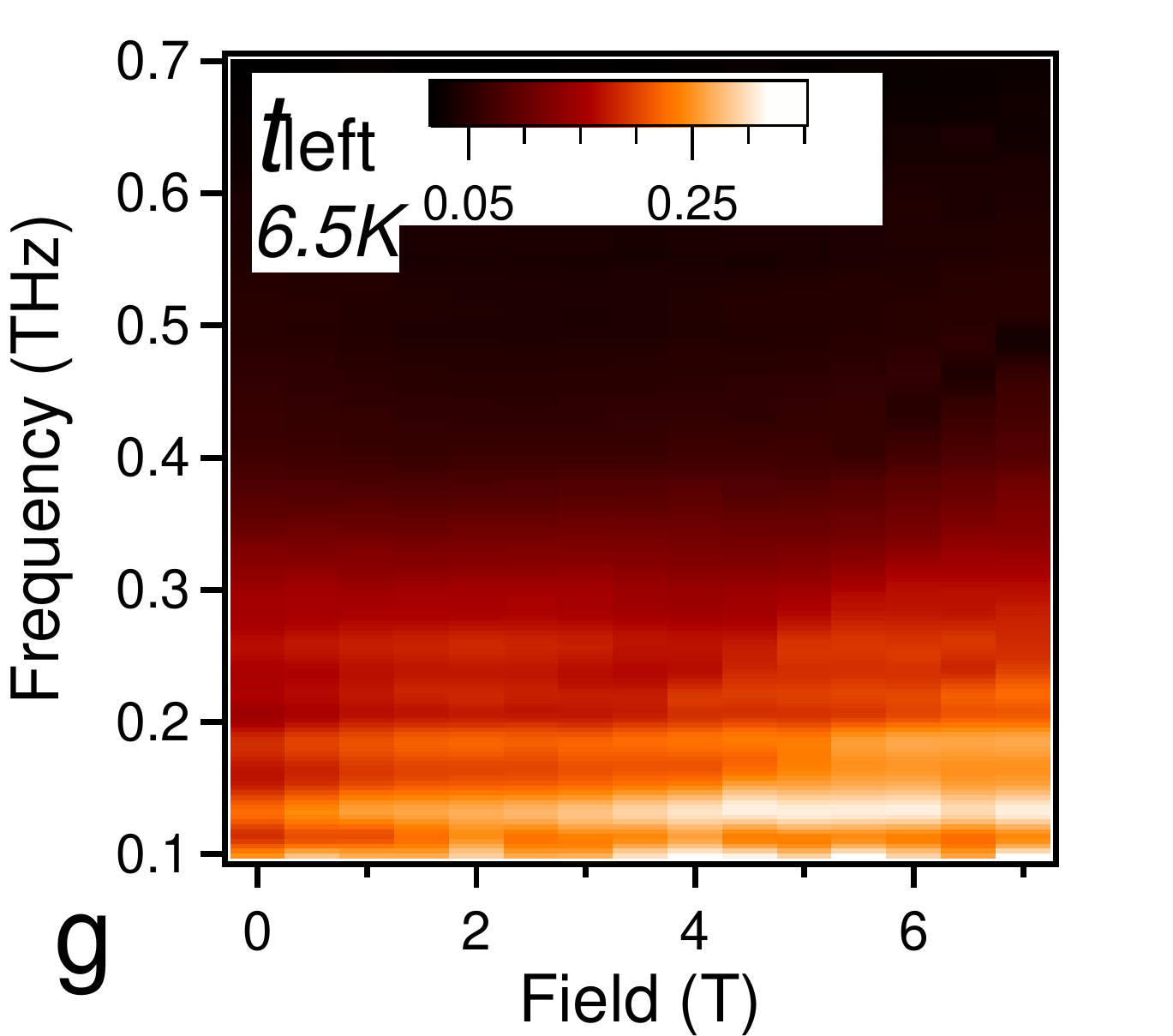}
\includegraphics[trim = 10 5 5 5,width=4.2cm]{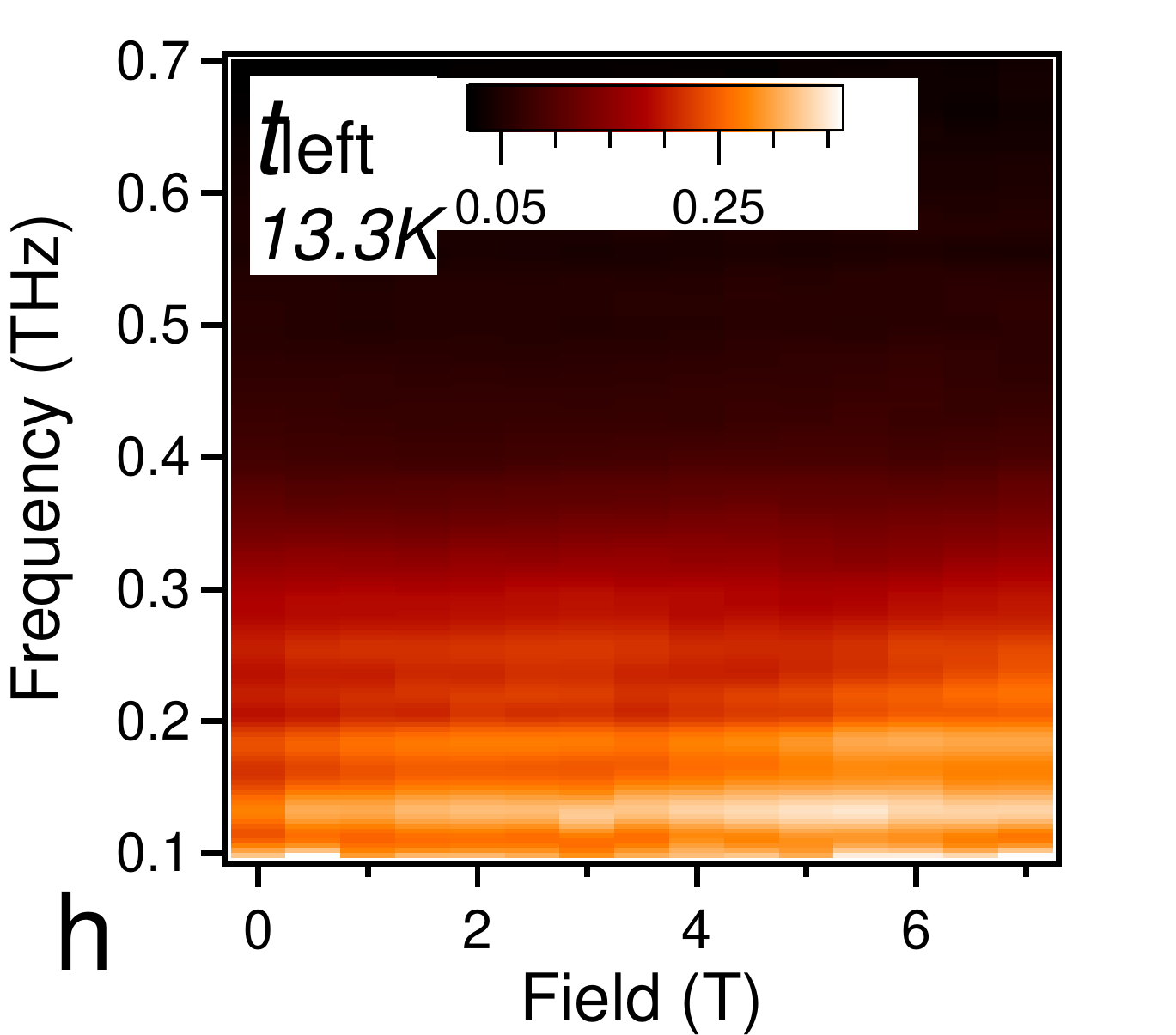}
\includegraphics[trim = 5 5 5 5,width=13.5cm]{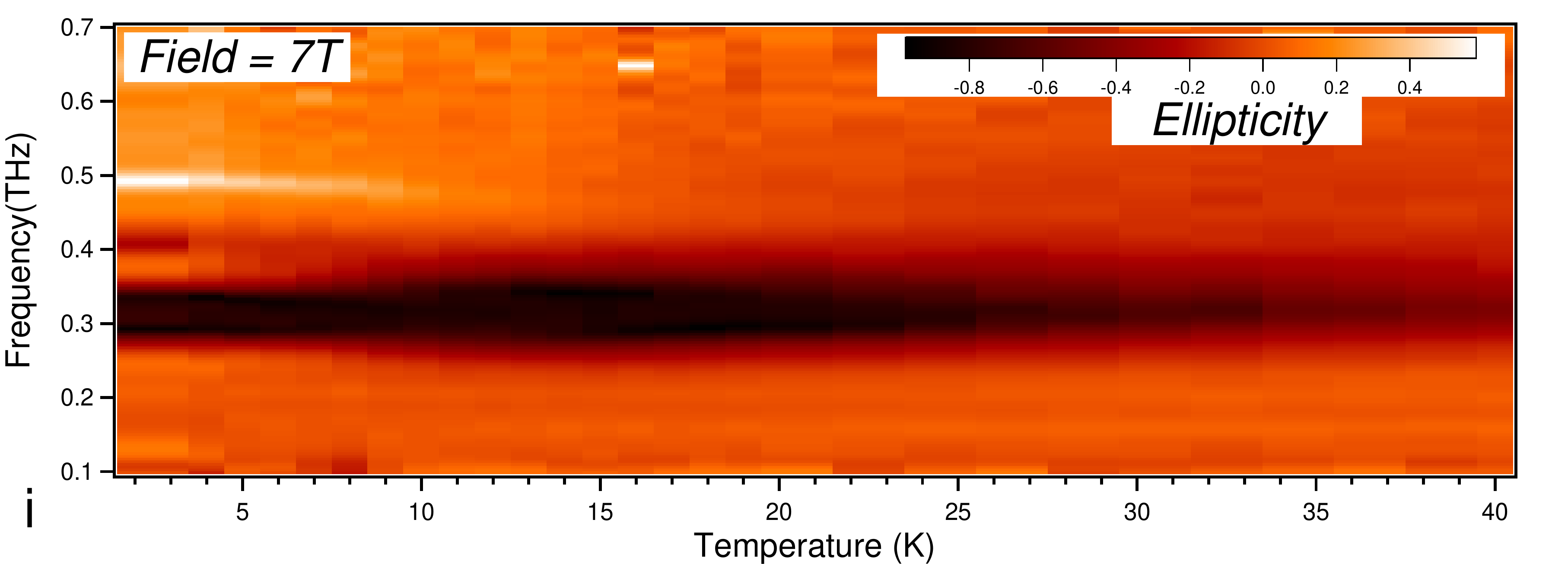}

\label{figSI2} 
\caption{ \textsf{\textbf{Transmission as a function of field and frequency in the Faraday geometry}}  (a)-(h) Transmission of the right and left circular polarized channel measured from sample A at different temperatures from 1.65 K to 13.3 K plotted as a function of frequency and field. Color scales are the same for RCP and LCP with the same temperature. (i) Ellipticity as a function of frequency and temperature obtained with $\eta = (\mathit{t_{right}-t_{left}})/(\mathit{t_{right}+t_{left}})$, with an applied field of 7 T. Data were taken at T = 1.65 K; 4.3 K, 5 to 20 K in 1 K step and then 22-40 K in 2 K step.}
\end{figure*}

\begin{figure*}
\includegraphics[trim = 10 5 5 5,width=5.cm]{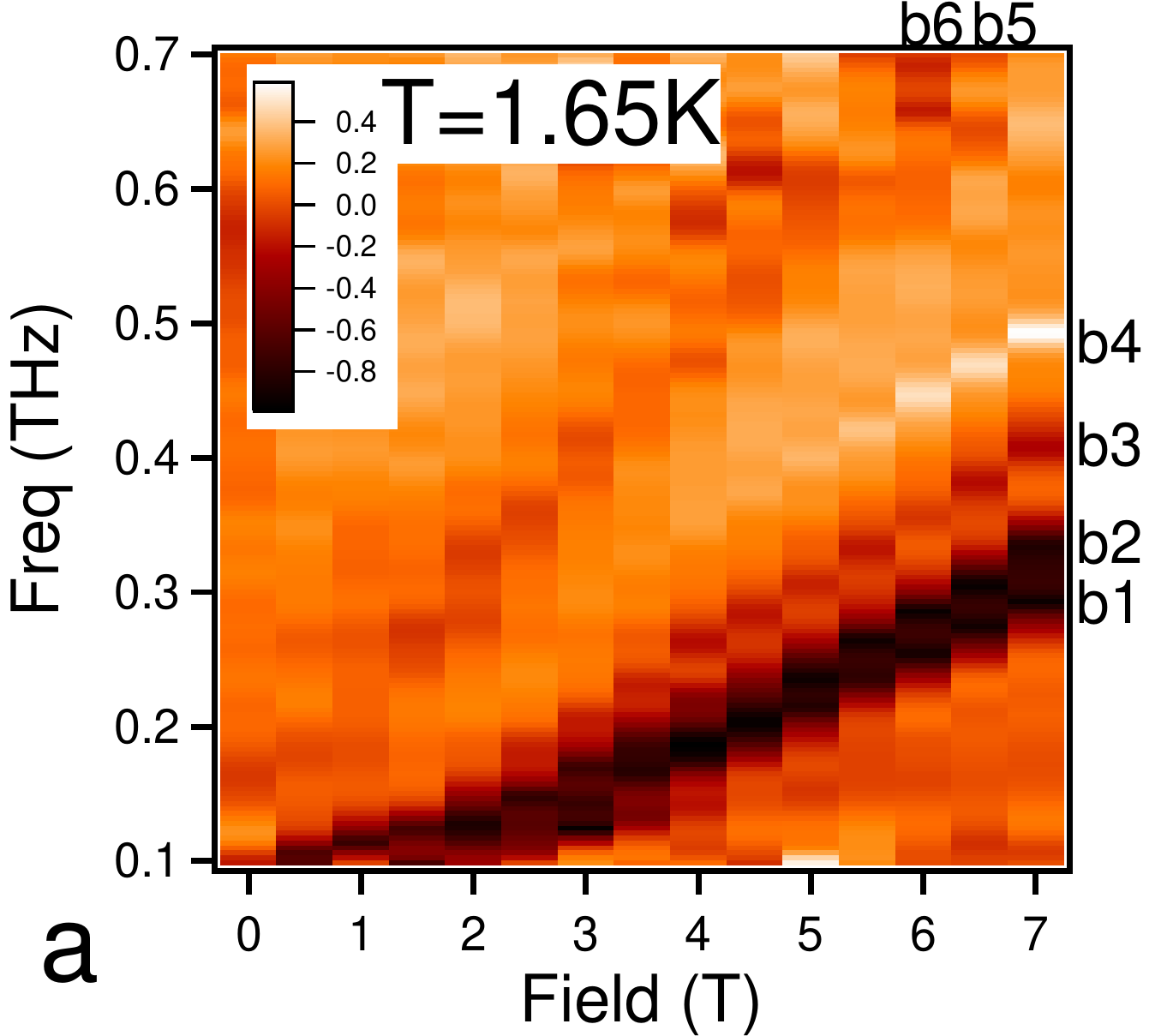}
\includegraphics[trim = 10 5 5 5,width=5.cm]{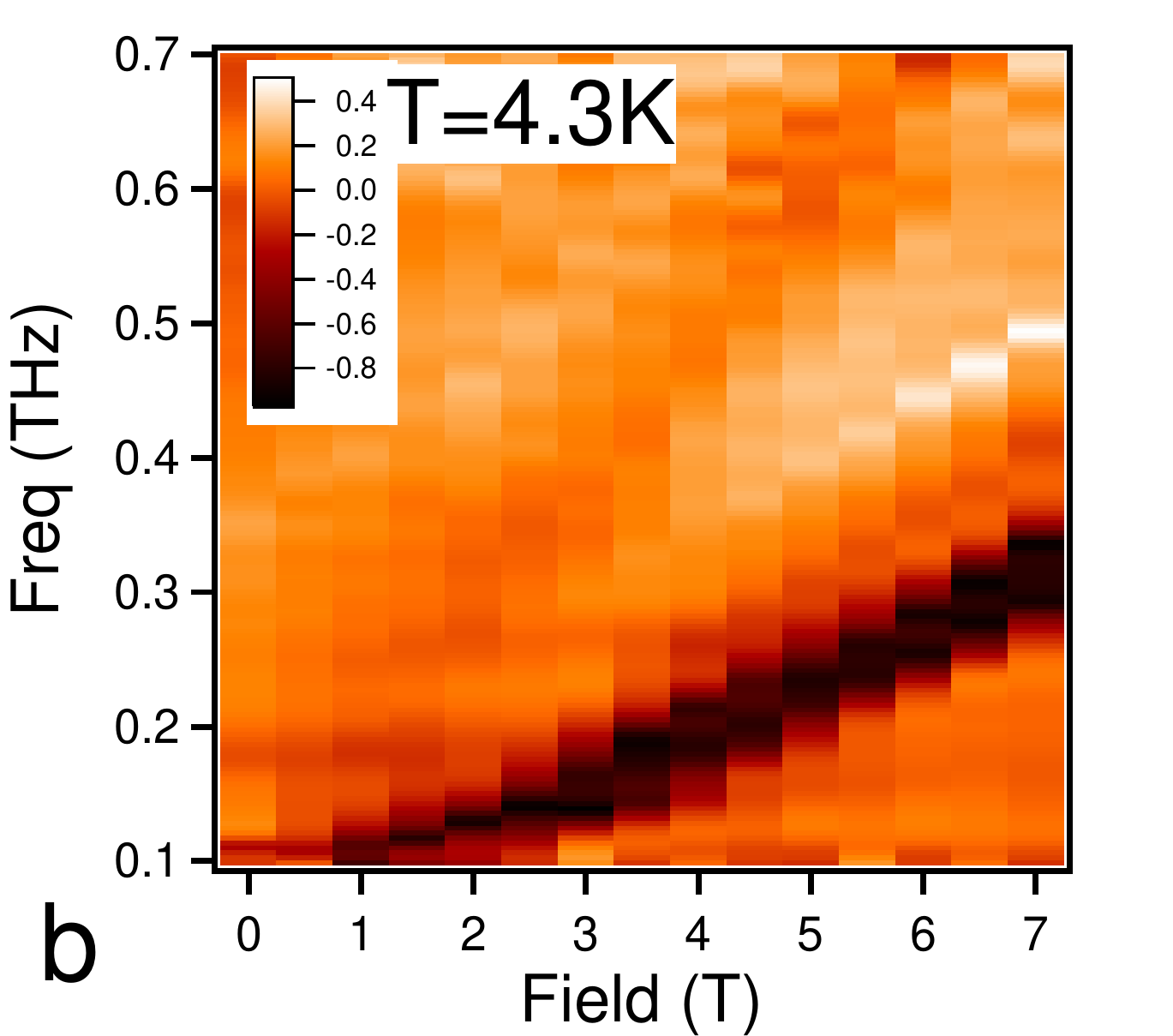}
\includegraphics[trim = 10 5 5 5,width=5.cm]{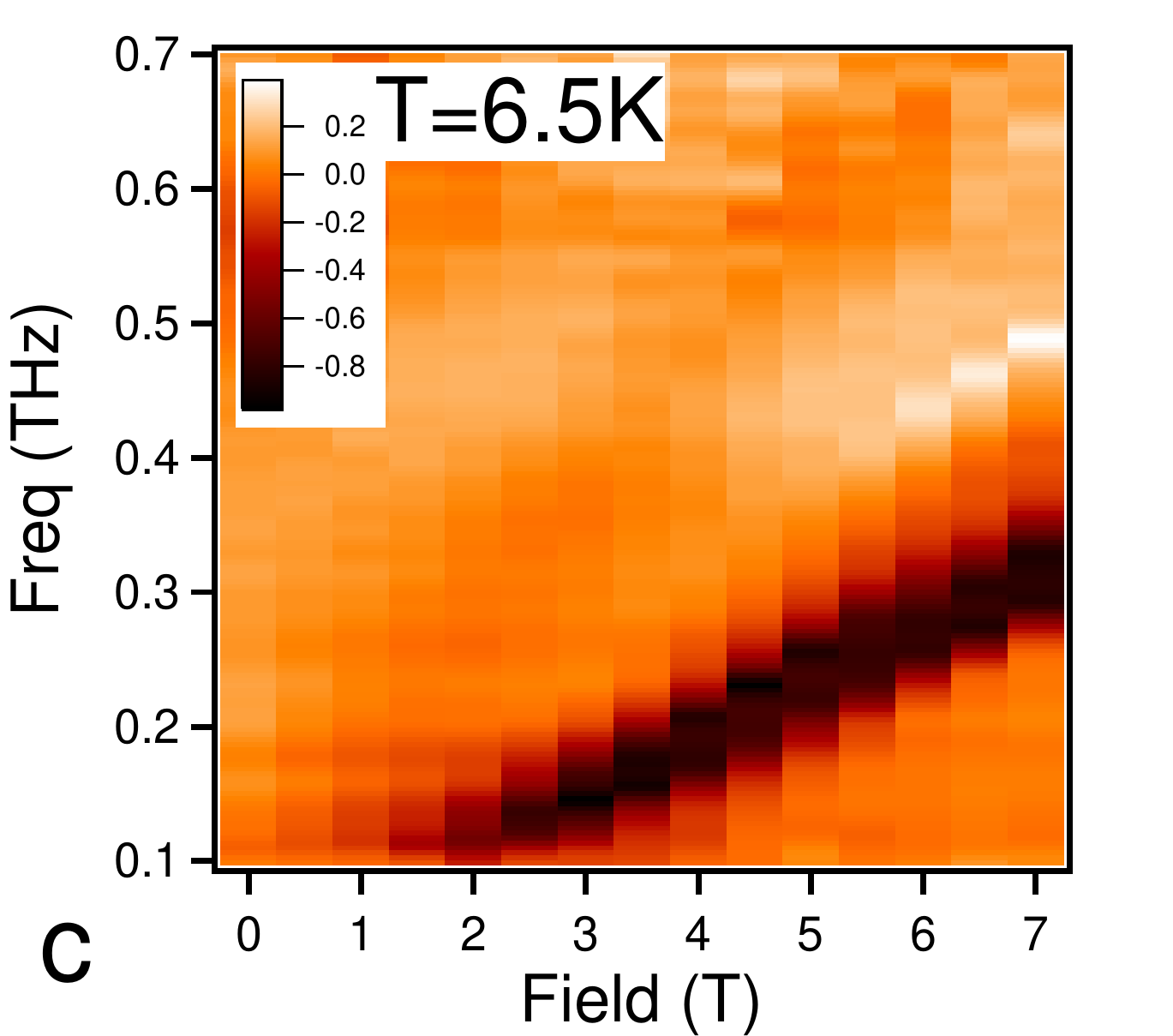}
\includegraphics[trim = 10 5 5 5,width=5.cm]{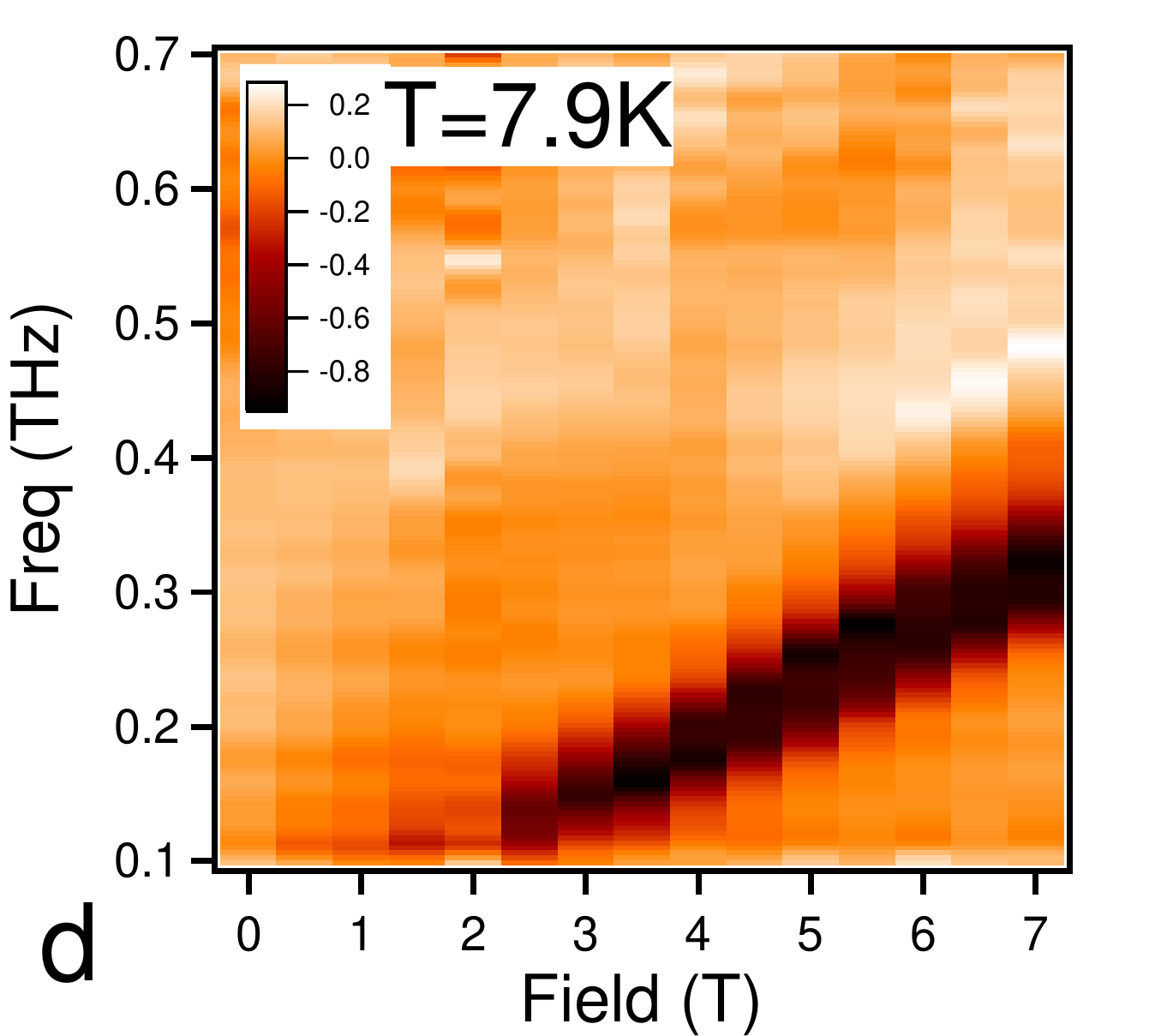}
\includegraphics[trim = 10 5 5 5,width=5.cm]{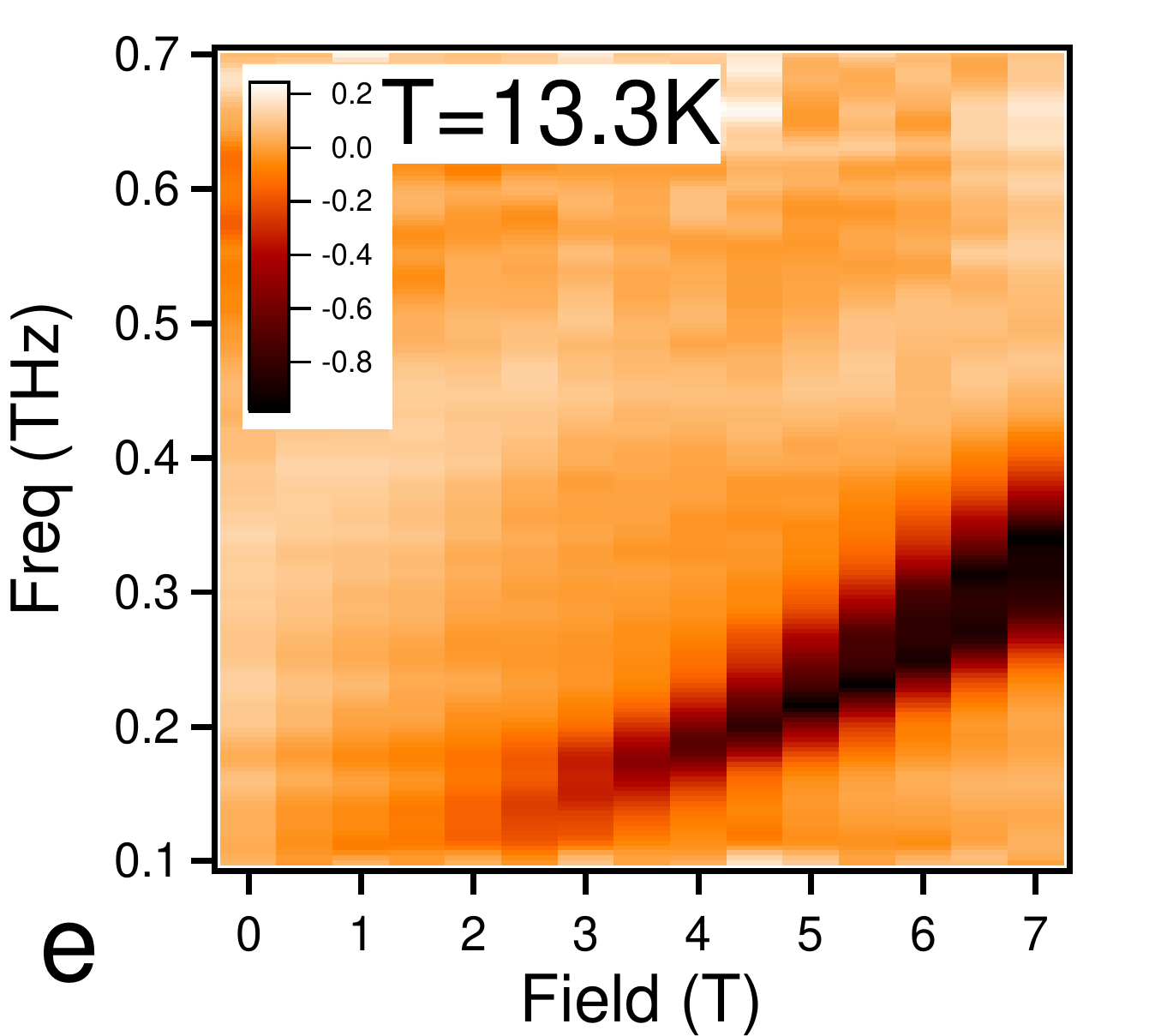}
\includegraphics[trim = 10 5 5 5,width=5.cm]{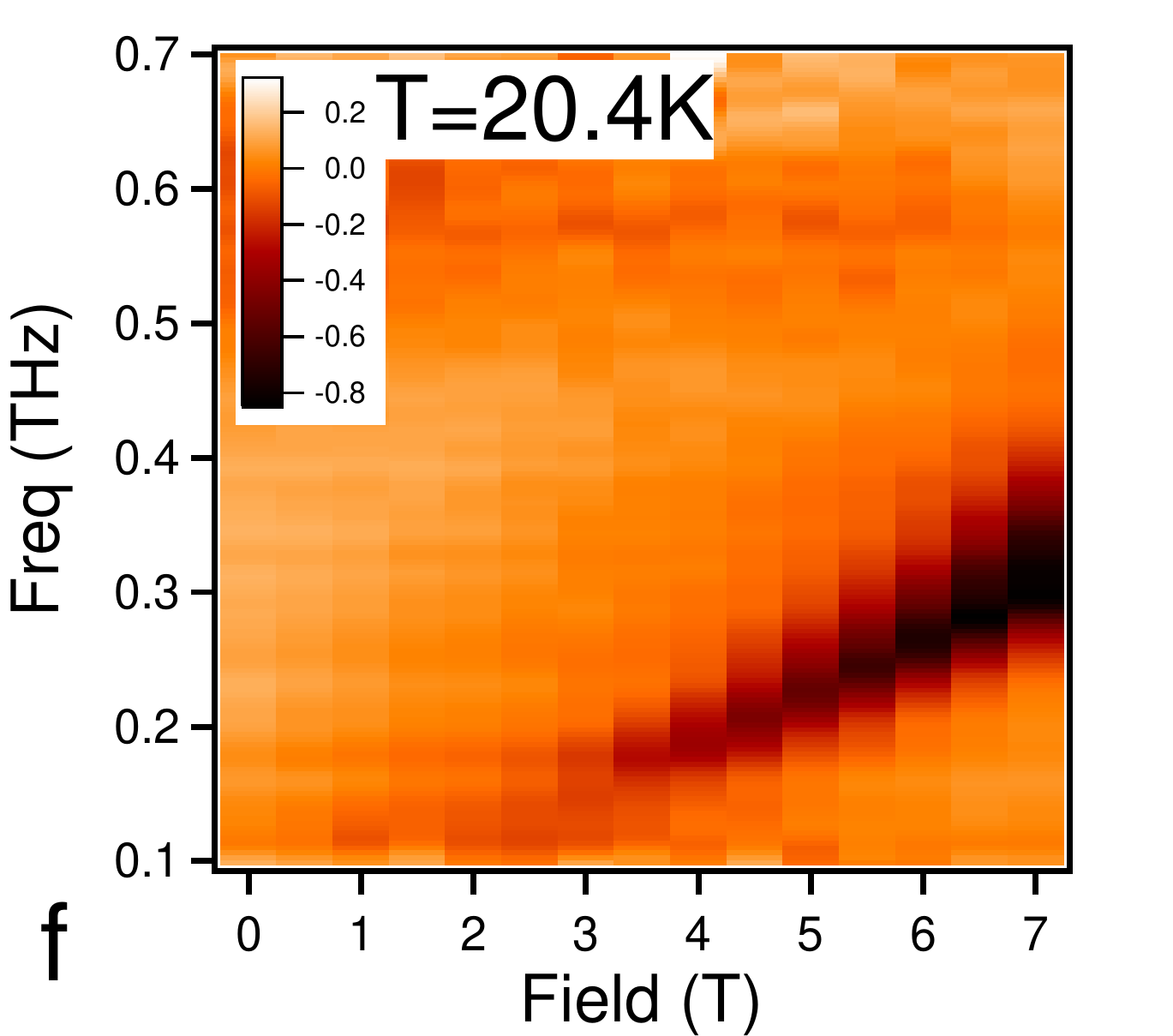}
\label{SI2} 
\caption{ \textsf{\textbf{Ellipticity from Faraday geometry measurement}}  Ellipticity measured from sample A at temperatures from 1.65 K to 20.4 K.  Different signs of the ellipticity indicates absorptions in different circular channels.}
\end{figure*}

We present in Fig. SI2(a) to (h) the frequency and field dependence of the transmission magnitude of the right and left circular polarized light at temperatures from 1.65 K to 13.3 K, measured from Sample A. As Sample A is thicker than Sample B, data from A show stronger absorption for \textit{b1} and \textit{b2}. However, the lower transmission from A also lowers the contrast between the magnetic excitation and the background, thus diminishes the visibility of \textit{b5} and \textit{b6}. Several features are evident from the data: 1) as temperature increases, all absorptions widen in frequency and lose intensities; 2) the slopes of those excitations have no obvious temperature dependence in the frequency - field plot; 3) \textit{b3}, \textit{b5} and \textit{b6} disappear at a temperature around 8 K; 4) \textit{b4} disappears at temperature between 8 and 13 K; 5) \textit{b1} and \textit{b2} widen and merge together as temperature increases; 6) the spectra at 1.65 K and 4.3 K appear almost identical, except that at 1.65 K the absorptions are slightly sharper and stronger; 7) the higher frequency branches \textit{b5, b6} broaden significantly at low field; and 8) for the frequency range below 0.25 THz, the application of magnetic field increases the transmission in both the right and left channel, an effect which reduces as temperature is increased.  To demonstrate the temperature dependence of the magnetic excitations in detail, measurements on Sample A are shown as a function of temperature while holding the applied field constant at 7 T. Fig. SI2(i) shows the ellipticity $\eta$ of the transmitted pulse at 7 T as calculated from $\eta = (\mathit{t_{right}-t_{left}})/(\mathit{t_{right}+t_{left}})$, from 1.65 K to 40 K. The results show similar trends as those observed from the plots shown in Fig. 3(a) to (h).  The temperature dependence is consistent with the accepted scale of the exchange constants$^{14}$; Fig. SI2(i) shows that the single ion regime is reached above 12 K ($\approx 6$ $J_{zz}$).


Excitations \textit{b1} and \textit{b2} are the strongest, and stay separated until around 25 K, above which temperature the magnetic absorption becomes considerably weaker. On the other hand, \textit{b3} and \textit{b4} disappear at temperatures as low as 8 and 10 K, respectively. The resonant frequency of all the excitations show insignificant temperature dependence; except for \textit{b3}and \textit{b4} which showed a slight softening before disappearing.

As one can see from the main text, systematic errors in TDTS show up as horizontal lines in the intensity plots of the field dependent transmission data (e.g. Fig. SI2 (a) to (h)). Those type of systematic modulations of the data can be significantly reduced by plotting the ellipticity instead of the transmission amplitude. From the definition $\eta = (\mathit{t_{right}-t_{left}})/(\mathit{t_{right}+t_{left}})$, it is clear that the common error in the measurement will be canceled out in computing $\eta$. Thus it is a very useful quantity to look at when studying the subtle trends of the data, as shown in Fig. SI3.

\subsection{Field-Induced Transparency}

\begin{figure*}
\includegraphics[trim = 10 5 0 5,width=10.5cm]{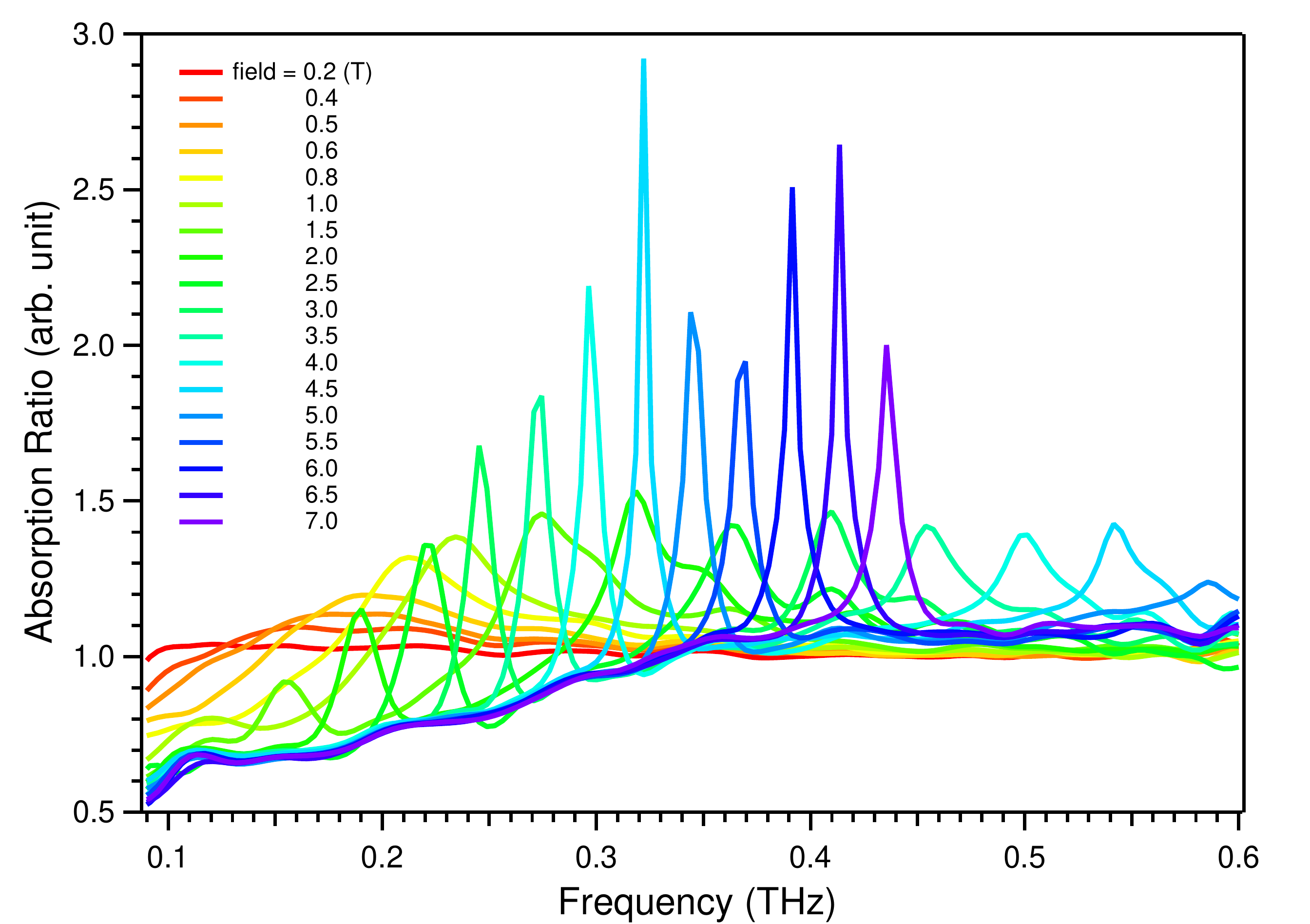}
\label{SI1} 
\caption{ \textsf{\textbf{Absorption ratio from Voigt geometry measurement}} Absorption as a function of frequency at different applied fields from Voigt geometry measurement with the electric field polarization along the \textbf{x} direction, divided by the corresponding zero field spectrum, measured from sample A at 1.65 K. Higher relative value indicates stronger absorption at a given field, as compared with the zero field measurement. The two groups of peaks correspond to \textit{c2} and \textit{c4}, respectively. }
\end{figure*}

\begin{figure*}
\includegraphics[trim = 10 5 0 5,width=8.5cm]{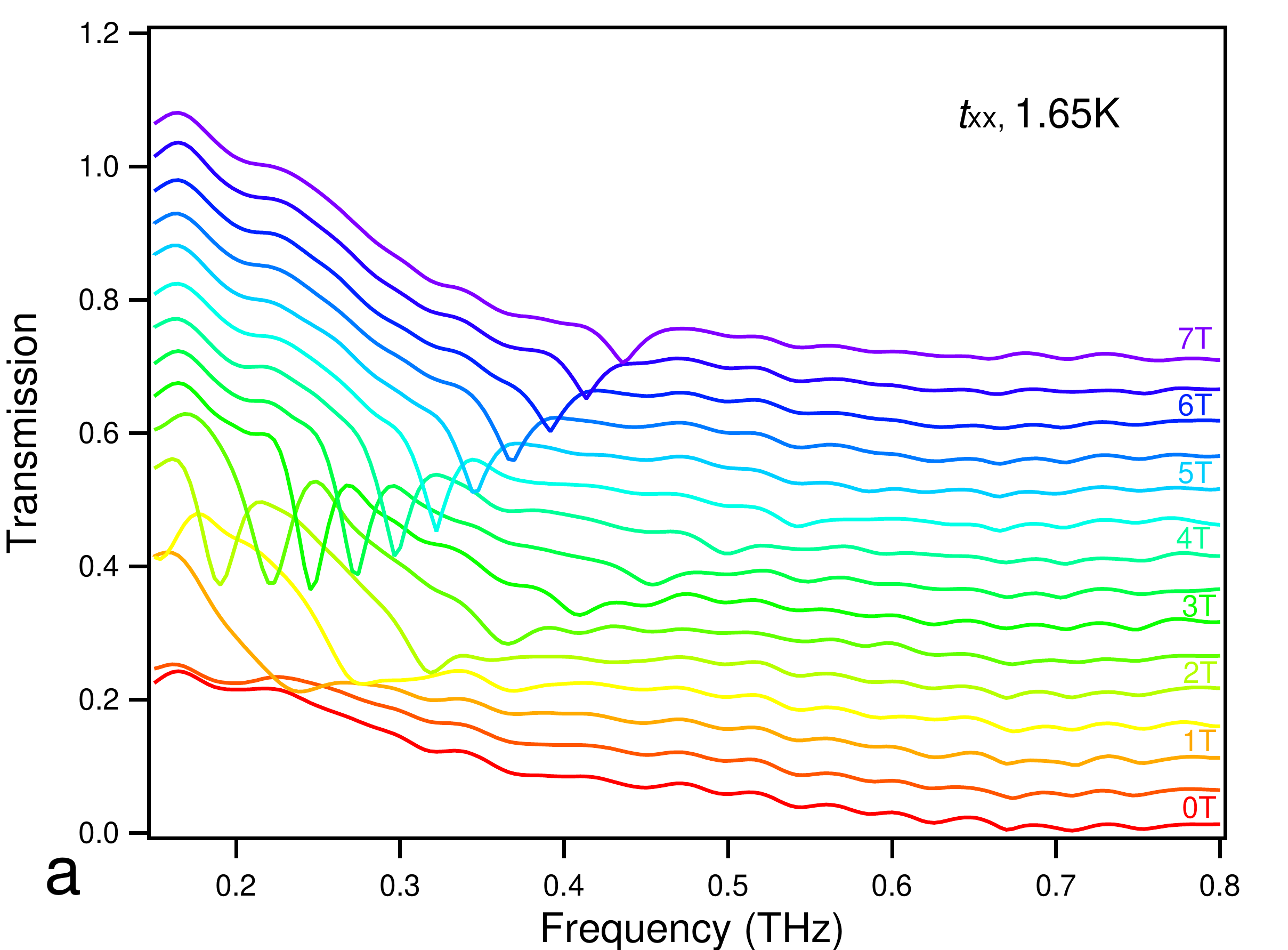}
\includegraphics[trim = 10 5 0 5,width=8.5cm]{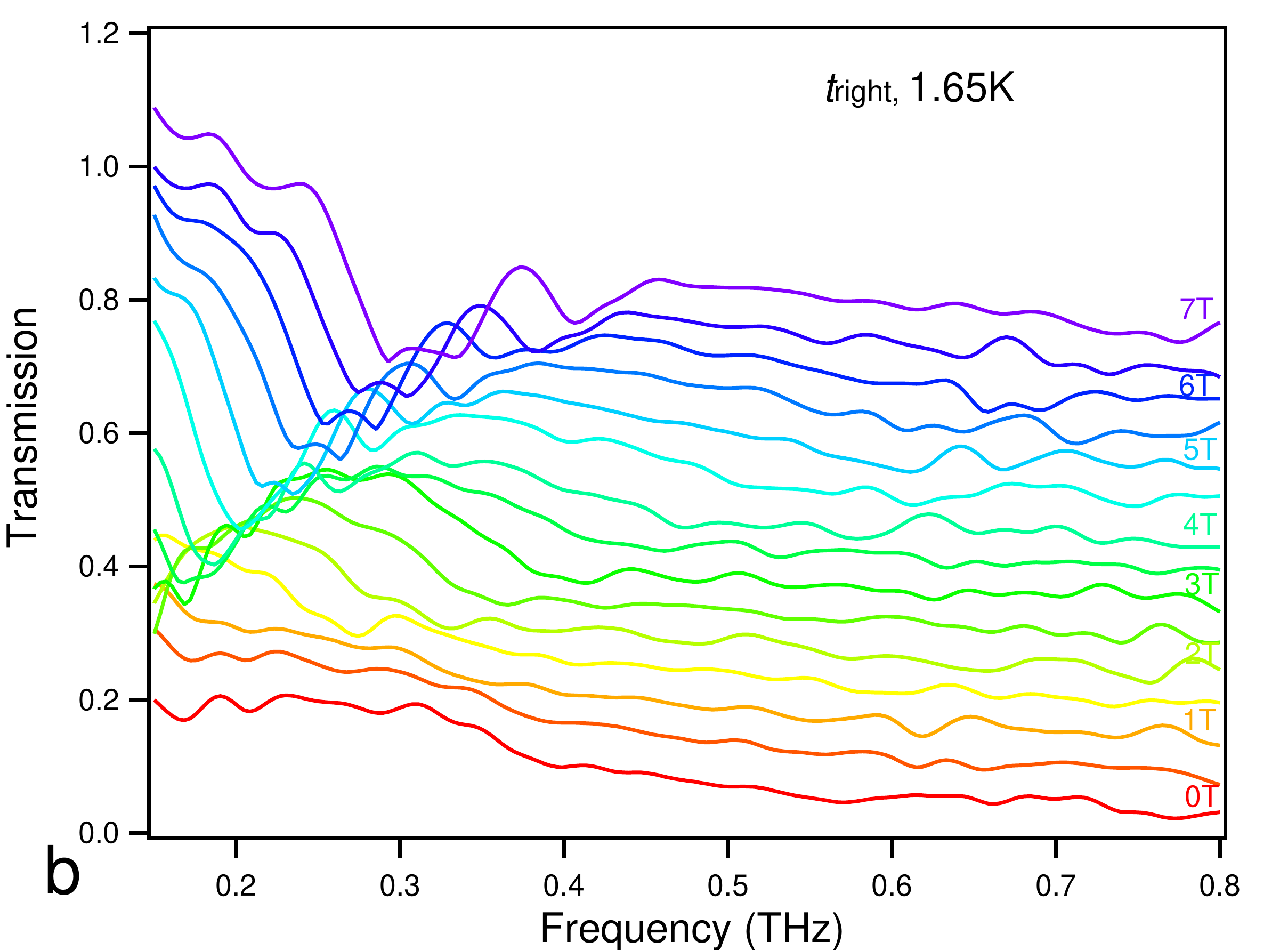}
\label{SI2} 
\caption{ \textsf{\textbf{Transmission spectra at 1.65 K}}  Transmission spectra as a function of frequency measured at 1.65 K from sample A with (a) t$_{xx}$ from the Voigt geometry and (b) RCP from the Faraday geometry, the spectra are shifted in the vertical direction for clarity.}
\end{figure*}

The field induced transparency feature presents itself in the form of increased transmission in the frequency ranges away from magnetic excitations as an external magnetic field is applied (most noticeably below 0.25 THz). The observation of such effects indicates the existence of anomalous magnetic absorptions at low fields, which in our case, hints at the existence of string-like excitations.

\begin{figure*}
\includegraphics[trim = 10 5 0 5,width=10.5cm]{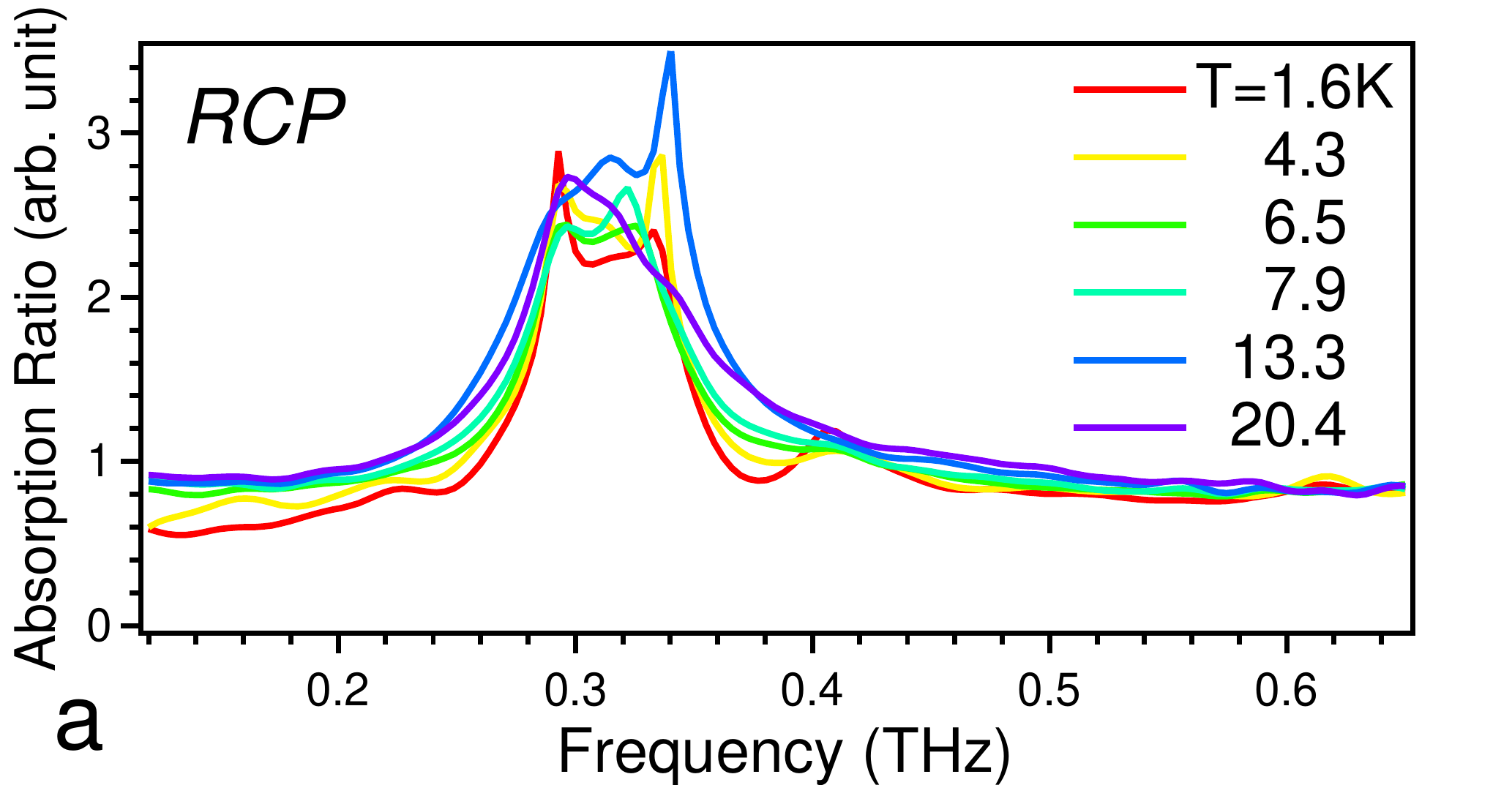}
\includegraphics[trim = 10 5 0 5,width=10.5cm]{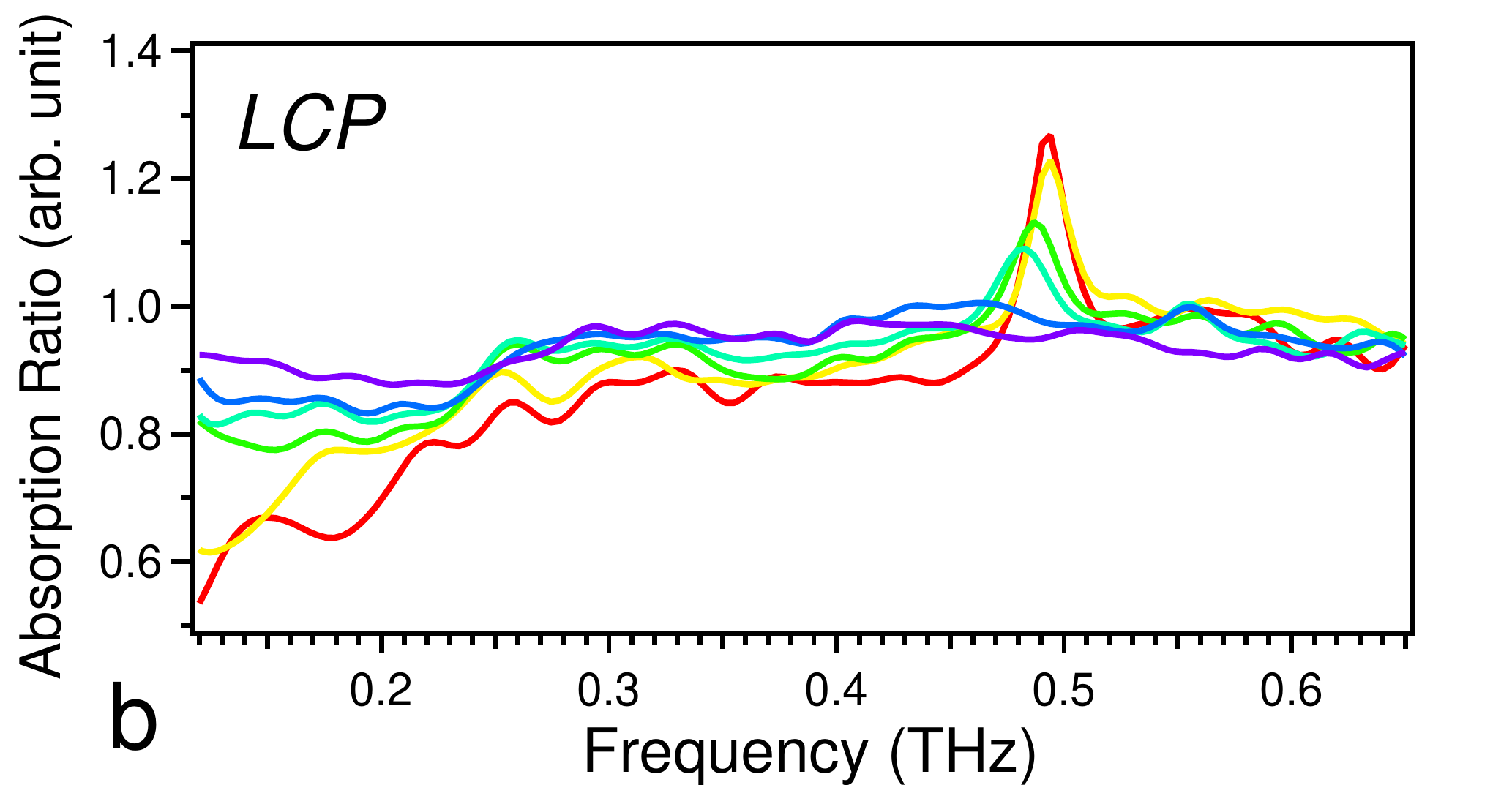}
\label{SI2} 
\caption{ \textsf{\textbf{Absorption ratio from Faraday geometry measurement}}  Absorption with 7 T applied field as a function of frequency at different temperatures from Faraday geometry measurement with (a) RCP and (b) LCP, divided by the corresponding zero field spectrum, measured from sample A.}
\end{figure*}

In Fig. SI4, we present the data taken from sample A at 1.65 K in the Voigt geometry, plotted in the form of ratio of absorption under a fixed magnetic field ($-\ln{t(H)}$) with that obtained from zero field scans ($-\ln{t(H=0)}$). Plotting the data in this fashion would reduce the spectroscopic features due to systematic noise in the measurement, thus makes it easier to observe the subtle trends. Since ($-\ln{t(\omega)}$) is proportional to the absorption, a value larger than unity of this quantity would mean at that particular frequency, the transmission in field is lower than zero field. On the other hand, when the absorption ratio is lower than unity, there is a field induced transparency, as clearly shown in the data in Fig. SI4.  In this case, the sample becomes more transparent as the field is increased and magnetic absorptions move to higher energy.  It is also interesting to notice that, in the two branches visible in Fig. SI4 (\textit{c2} and \textit{c4}), the width of \textit{c2} is relatively constant, while \textit{c4} broadens significantly as field is decreased. The original transmission spectra as a function of frequency at 1.65 K at different fields for the t$_{xx}$ in the Voigt geometry and RCP in the Faraday geometry are shown in Fig. SI5. 

To understand the low frequency absorption better, we look at the temperature dependence in Fig. SI6, with data taken from the Faraday geometry. Shown in the figure is the absorption ratio for RCP and LCP between scans taken at 7 T and zero field at temperatures from 1.65 K to 20.4 K. It is clear from the data that the field induced transparency reduces as temperature increases, and is almost completely absent for temperature higher than 20 K, about the same temperature thermal fluctuations destroy spin correlations.

\subsection{Spin Wave Calculation}

The spin wave calculations take into account the Zeeman term with an anisotropic g tensor and four possible bilinear spin interactions as discussed in Ref. 14.  The classical spin wave modes are obtained in the linear spin wave theory by expanding the Hamiltonian in the large $S$ limit about one of the classical ground states, which is numerically determined. We use the interaction parameters as obtained from the recent fits to INS experiment on Yb$_2$Ti$_2$O$_7$ as the starting values$^{14}$. From the detailed field dependent scans and high signal to noise in TDTS, we can refine the values for the exchange parameters. When fitting to the magnetic absorptions in the Voigt geometry above 3 T with the method of least squares, we find a best fit in units of meV of $J_1 = -0.08$, $J_2 = -0.26$, $J_3 = -0.25$, and $J_4 = -0.02$ within the parameter search intervals which are twice as large as the uncertainty intervals in Ref. 14, which correspond to $J_{zz} = 0.16$, $J_{\pm} = 0.065$, $J_{\pm\pm} = 0.02$ and $J_{z\pm} = -0.134$.

There are a total of four non-degenerate spin wave modes, but two of these are predicted to have vanishing spectral weight as $q \rightarrow 0$ in the Faraday geometry.   With 4 spins in the unit cell, there are 8 transverse spin components. Their harmonic interactions have the symmetry of the dihedral group $D_{4h}$, including two $C_4$ and three $C_2$ proper rotations (group $D_4$) times the inversion.  These 8 transverse spin components contain all four 1-dimensional irreducible representations ($A_{1u}$, $A_{2u}$, $B_{1u}$, $B_{2u}$) and two copies of the two-dimensional irreducible representations $E_g$. Since the magnetic field components $H_x$ and $H_y$ transform as $E_g$, only the $E_g$ modes are expected to be visible in optics in the Faraday geometry.  The visible $E_g$ doublets have the highest and lowest frequencies.  The modes invisible in the Faraday geometry are $A_{1u}$ + $A_{2u}$ and $B_{1u}$ + $B_{2u}$ with the former slightly higher than the latter for coupling constants of Ref. 14.   The splitting of the visible $E_g$ modes is field-independent and equal to $4 |J_1+J_2| S$  = 0.62 meV = 0.15 THz with the same coupling constants. The visibility of these modes can be understood heuristically in that one can see that in all four modes individual spins precess elliptically around the applied $<$001$>$ field, with a large right-hand component and a small left-hand component.   In the lowest-frequency mode the right-hand components add up in phase. In the highest-frequency mode they cancel out, leaving a small left-hand component. In the intermediate modes right- and left-hand components cancel out, leaving only a longitudinal oscillation.

The spin wave calculation results bear some resemblance and some notable difference with our TDTS results.   We see a lower multiplet of 4 states that presumably corresponds to magnon-like (e.g. strings of length one) excitations.  Note that all of them exhibit about the same $g$-factor of approximately 3.2 as predicted.  The spin-wave result also explains the otherwise anomalous appearance of the LCP $b4$ polarized absorption in a $\bf{+z}$  $<$001$>$ directed magnetic field.   But there are also notable differences.   In the lowest multiplet of states we see all 4 instead of 2 excitations predicted by theory.  We also see two excitations ($b5$ and $b6$) in another group of states at higher energy with approximately the twice the $g$-factor of the lower multiplet.   If the lower multiplet can be understood as spin-wave like excitations, then we can understand these higher energy excitations in the high field regime as two-magnon states or strings of approximate length two.  

\subsection{Two-Magnon Calculation}

\begin{figure*}
\includegraphics[trim = 5 5 5 5,width=8cm]{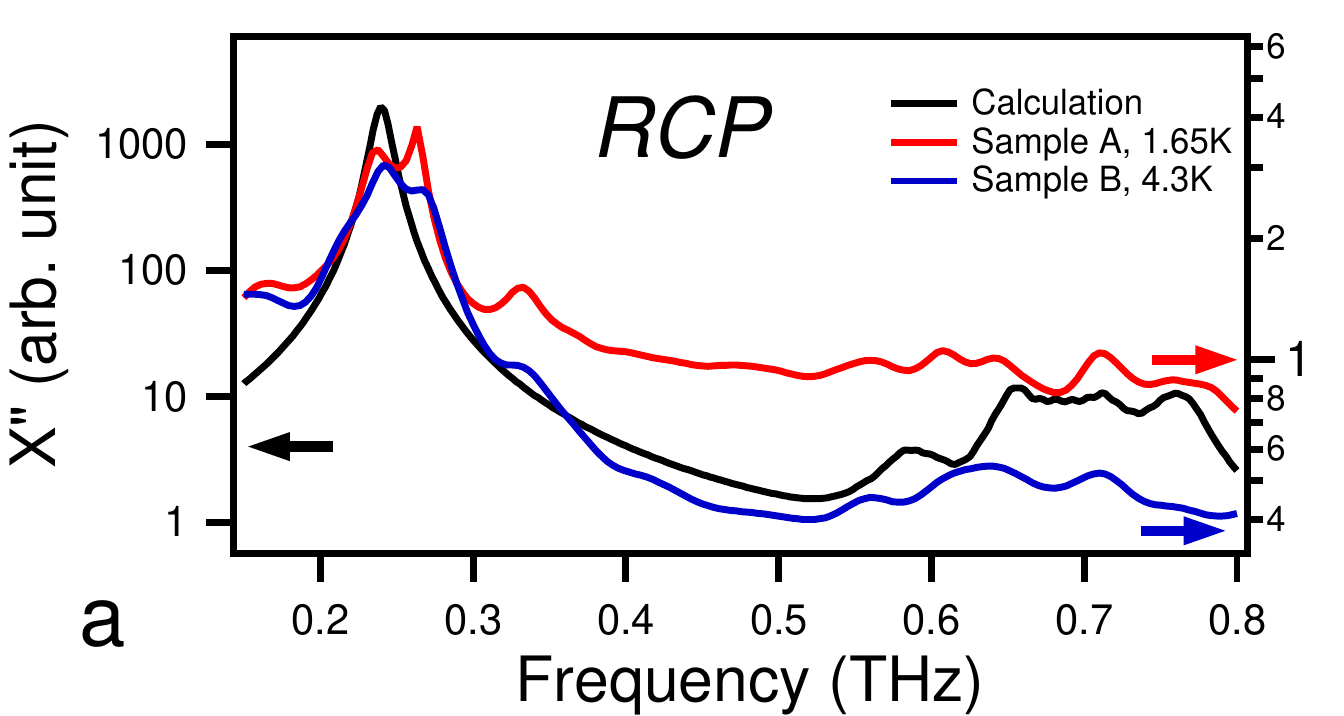}
\includegraphics[trim = 5 5 5 5,width=8cm]{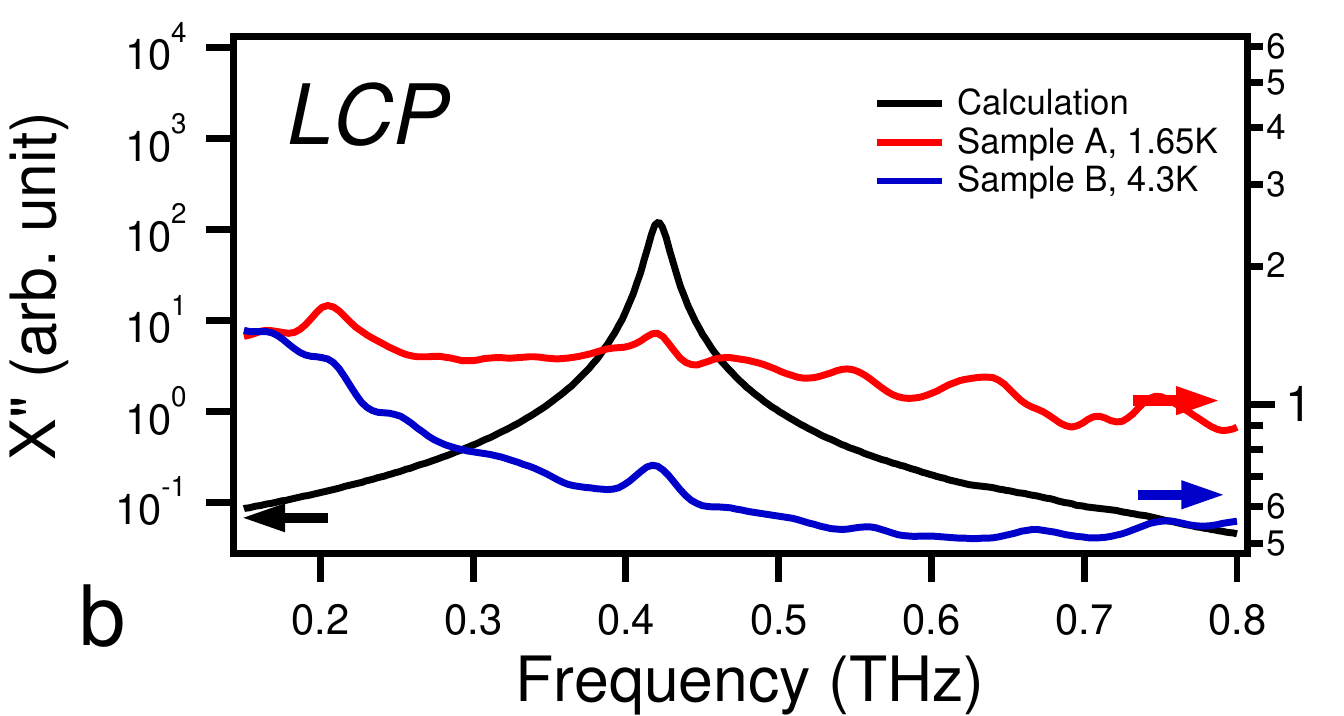}
\includegraphics[trim = 5 5 5 5,width=8cm]{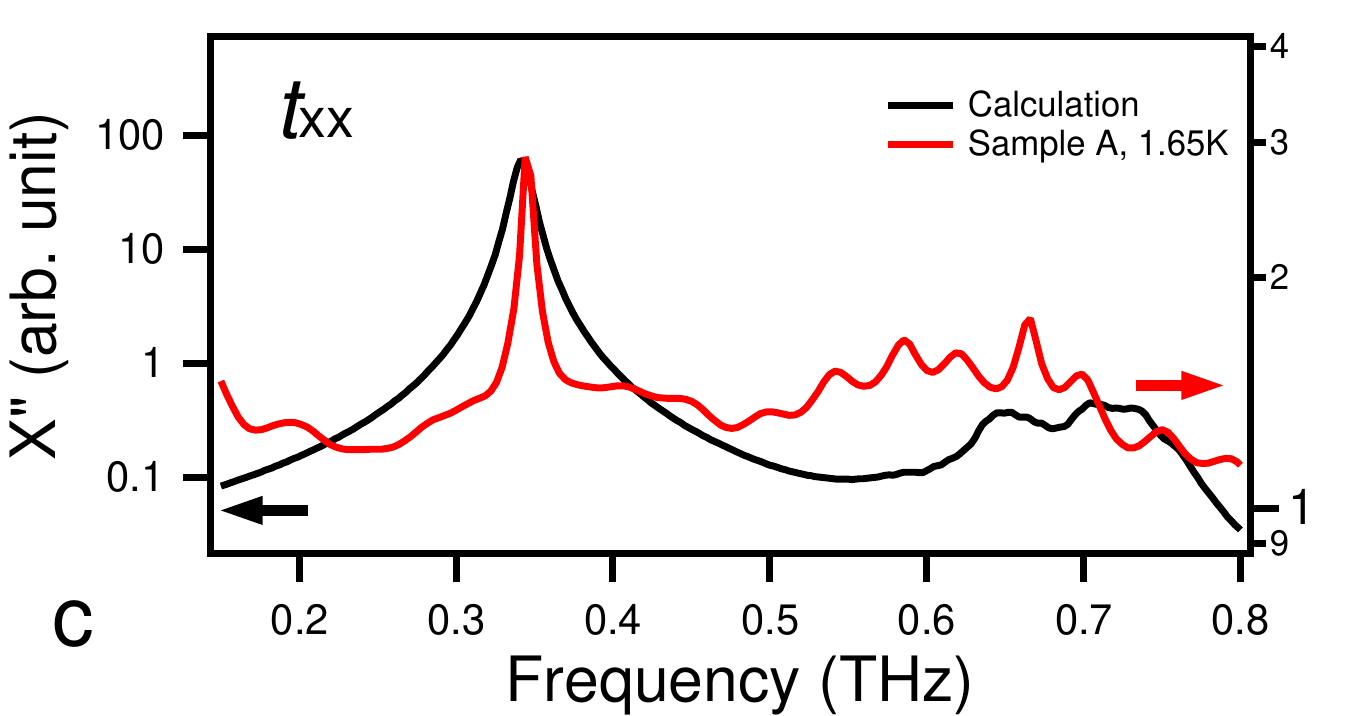}
\includegraphics[trim = 5 5 5 5,width=8cm]{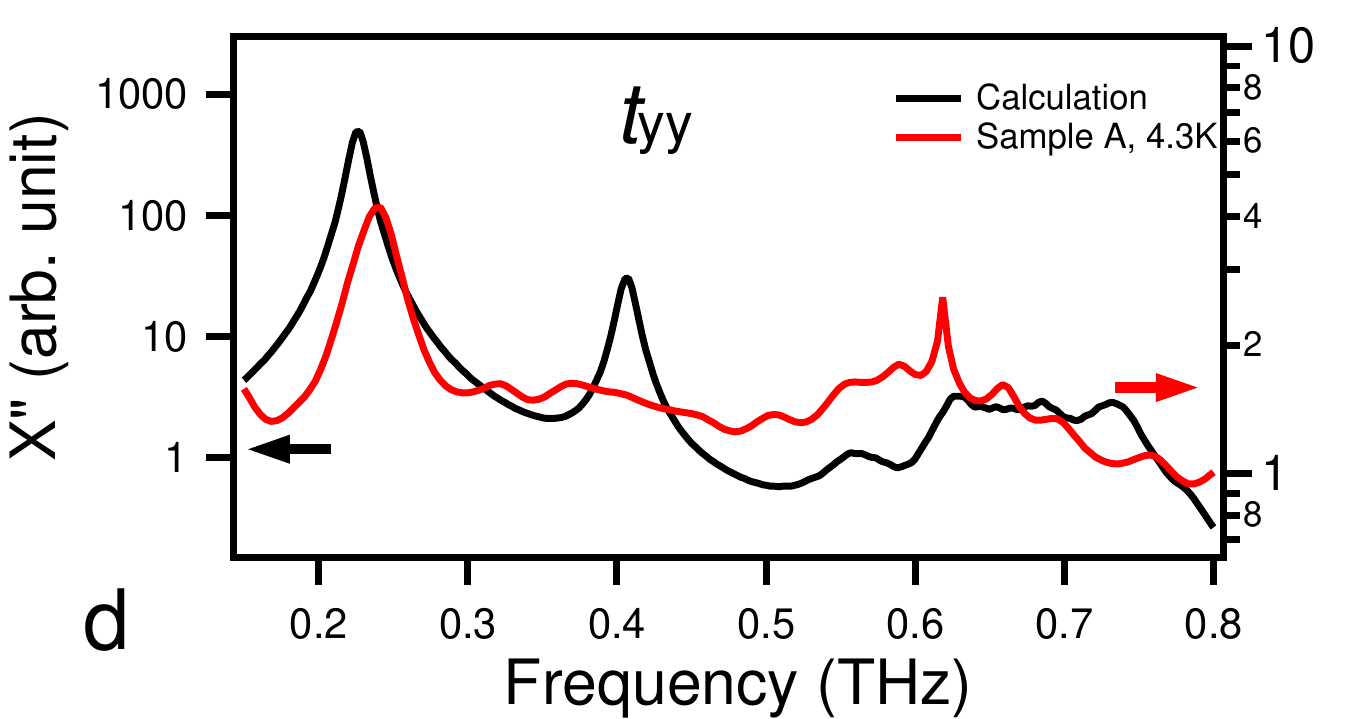}
\label{SI6} 
\caption{ \textsf{\textbf{Theoretical and experimental spectra of dynamic susceptibility}}  Dynamic susceptibility obtained from the theoretical calculation as explained in the text as well as from the measurement with (a) RCP and (b) LCP with an applied field of 5.27 T, and (c) t$_{xx}$ and (d) t$_{yy}$ with an applied field of 4.96 T.}
\end{figure*}

As an attempt to understand the origin of the two-magnon peaks, we calculated the spectral function of the two-magnon continuum. To do this, we rewrote the Hamiltonian in a set of locally rotated axes such that at a given site the spin in the classical ground state points along the local z-direction. From this one can readily write an effective Hamiltonian $H_{eff}$ in the subspace of one and/or two magnons at zero total momentum. $H_{eff}$ has nearest-neighbor magnon hopping (deriving from terms like $S_{i}^{x}S_{j}^{x}$ and $S_{i}^{x}S_{j}^{y}$ in the full Hamiltonian), nearest-neighbor magnon interaction (deriving from the $S_{i}^{z}S_{j}^{z}$ terms) and single-magnon to two-magnon tunelling terms (deriving from terms like $S_{i}^{x}S_{j}^{z}$ and $S_{i}^{y}S_{j}^{z}$). The spectral function of this Hamiltonian (with the optimal parameters obtained from linear spin wave theory mentioned in Section E) was obtained using Lanczos diagonalization. Convergence of the spectral function was checked by varying both the system size $L$ (achieved at $L=10$) and the number of Lanczos iterations $\Lambda$ (achieved at $\Lambda=50$).

At a given external magnetic field, four spectral functions were obtained corresponding to the RCP and the LCP channels in Faraday geometry and the X and the Y channels in Voigt geometry. Each spectral function has sharp peaks corresponding to the single-magnon excitations followed by a two-magnon continuum. The single-magnon energies thus obtained were found to differ somewhat from the classical spin-wave results for the lowest branch. This difference was larger at low fields and became smaller as the field was increased ($10\%$ at 3 T to $3.5\%$ at 7 T). The peaks in the two-magnon continuum region were not found to agree well with the experimental peaks (except perhaps for the second peak in the RCP channel). 

A comparison between the calculations and the experimental data is shown for one representative field in Fig. SI7. Here the parameter being plotted is the imaginary part of the dynamic susceptibility $\chi^"(q \rightarrow 0, \omega)$. The experimental spectra are obtained from $-\ln{t(H)}$/$(\omega d)$, with \textit{d} being the thickness of the sample. Spectra obtained this way represent the dissipative part of material parameter, which in the particular case studied here, is proportional to $\chi^"$. Note at frequencies above 0.7 THz, absorption due to optical phonons become substantial, and causes additional contribution to the experimental spectra. As shown in the figure, the calculation provides qualitative description of the data.    

We also examined the variation in the single-magnon energies at $q \rightarrow 0$ as one moves away from the optimal point in the four dimensional $(J_1, J_2, J_3, J_4)$ parameter space, which is achieved from fitting described in Section F. To do this, we obtained the $4 \times 4$ matrix that connects a small change in the coupling constants $(\delta J_1, \delta J_2, \delta J_3, \delta J_4)$ to the corresponding change in the single magnon energies (ordered such that $E_1$ is the highest mode) $(\delta E_1, \delta E_2, \delta E_3, \delta E_4)$ at a given external field and then looked at its eigenvectors and eigenvalues. This matrix has an eigenvector $(1, -1.18, 8.22, 1.38)$ with a very small eigenvalue $-0.004$, which implies that the single magnon energies are not very sensitive to small changes of the coupling constants in that particular direction. This direction translates to $(\delta J_{zz}, \delta J_{\pm}, \delta J_{\pm\pm}, \delta J_{z\pm})=(-8.38, -1.3, -2.31, 1.57)$. This result was also found to be quite independent of the magnitude of the applied external field. Thus,  in the present calculation, the value of the $J_3$ (or equivalently, the $J_{zz}$) parameter is not very well determined from fitting the single-magnon energies to spin wave calculations. 

We wanted to check whether the insensitivity of the magnon energies along the direction listed above persists at finite momenta also. But the full Lanczos diagonalization was carried out only at $q \rightarrow 0$. So, here we will just quote what linear spin wave theory says at finite momenta. The linear model was found to predict a greater shift in the single-magnon energies with a change in $J_3$. More specifically, the vector $(\frac{\delta E_1}{\delta J_3}, \frac{\delta E_2}{\delta J_3}, \frac{\delta E_3}{\delta J_3}, \frac{\delta E_4}{\delta J_3})$ at $k=(0,0,0)$ was found to be $(0.03,0.02,-0.04,-0.08)$. The components of the corresponding vectors at $k=(1,1,1), (1,1,0)$ and $(0,0,1)$ differed by at most $\pm 0.01$ from these values, indicating that using the full magnon dispersions will not significantly improve the uncertainty in $J_3$.

\bigskip


\end{document}